\newif\ifsingle

\newif\ifproofs

\newif\ifFullVersion
\FullVersiontrue 

\ifsingle
\documentclass[12pt,draftclsnofoot, onecolumn]{IEEEtran}		
\else		
\documentclass[10pt,final, twocolumn]{IEEEtran}
\fi

\IEEEoverridecommandlockouts
\usepackage{balance}
\usepackage{amsmath}
\usepackage{amsthm}
\usepackage{amssymb}
\usepackage{algorithm}
\usepackage{algorithmic}
\usepackage[bookmarks,colorlinks]{hyperref}
\usepackage{cite}
\usepackage{cases}
\usepackage{graphicx}
\usepackage{subfigure} 
\usepackage{color}
\usepackage{enumerate}
\usepackage{xcolor}
\usepackage{acronym}

\newtheorem{thm}{Theorem}
\newtheorem{lem}{Lemma}
 
\newtheorem{corollary}{Corollary}

\newcommand{\TarVec}{\mathbf{a}}
\newcommand{\TarMat}{\mathbf{A}}
\newcommand{\LMMSEMat}{{\boldsymbol\Gamma}}
\newcommand{\CovMat}{{\boldsymbol\Sigma}}
\newcommand{\LMMSEMatW}{\tilde{\boldsymbol\Gamma}}

\acrodef{adc}[ADC]{analog-to-digital convertor}
\acrodef{bilimo}[BiLiMO]{bit-limited MIMO radar}
\acrodef{dft}[DFT]{discrete Fourier transform}
\acrodef{cs}[CS]{compressed sensing}
 
\ifsingle
\newcommand{\figWidth}{0.65\columnwidth}
\newcommand{\figHeight}{height=2.9in}

\newcommand{\includefig}[1]{\includegraphics[width = 0.75\columnwidth]{#1} 	\vspace{-0.2cm}}
\else
\newcommand{\figWidth}{2.5in}
\newcommand{\figHeight}{1.75in}
\newcommand{\includefig}[1]{\includegraphics[width = \columnwidth]{#1} 	\vspace{-0.2cm}}

\fi 

\DeclareMathOperator{\diag}{diag}
\DeclareMathOperator{\blkdiag}{blkdiag}
\DeclareMathOperator{\sign}{sign}
\DeclareMathOperator{\Tr}{Tr}

\DeclareMathOperator{\vect}{vec}
\DeclareMathOperator{\EMSE}{\epsilon}
\DeclareMathOperator{\LMMSE}{\epsilon_M}

\begin{document}
%
\title{BiLiMO: Bit-Limited MIMO Radar via Task-Based Quantization}
%
%
%

\author{Feng~Xi,~\IEEEmembership{Member,~IEEE,}
        Nir~Shlezinger,~\IEEEmembership{Member,~IEEE,}
        Yonina~C.~Eldar,~\IEEEmembership{Fellow,~IEEE,}
\thanks{F. Xi is with the Department
of EE, Nanjing University of Science and Technology, Nanjing 210094 China (email: xifeng@njust.edu.cn). 
N. Shlezinger is with the School of ECE, Ben-Gurion University of the Negev, Beer-Sheva, Israel (e-mail: nirshl@bgu.ac.il).
  Y. C. Eldar is with the Faculty of Math and CS, Weizmann Institute of Science, Rehovot, Israel (e-mail: yonina.eldar@weizmann.ac.il). 
This work  was supported in part by the Benoziyo Endowment Fund for the Advancement of Science, the	Estate of Olga Klein -- Astrachan, the European Union’s Horizon 2020 research and innovation program under grant No. 646804-ERC-COG-BNYQ, from the Israel Science Foundation under grant No. 0100101, from Futurewei Technologies, and by the National Science Foundation of China  under grant No. 61571228. }
\vspace{-1.0cm}
}

\maketitle
	\pagestyle{plain}
	\thispagestyle{plain}

\begin{abstract}
Recent years have witnessed growing interest in reduced cost radar systems operating with low power. Multiple-input multiple-output (MIMO) radar technology is known to achieve high performance sensing  by probing with multiple orthogonal waveforms. However, implementing a low cost low power MIMO radar is challenging. One of the reasons for this difficulty stems from the increased cost and power consumption  required by analog-to-digital convertors (ADCs) in acquiring the multiple waveforms at the radar receiver. In this work we study reduced cost MIMO radar receivers restricted to operate with low resolution ADCs. We design bit-limited MIMO radar (BiLiMO) receivers which are capable of accurately recovering their targets while operating under strict resolution constraints. This is achieved by applying an additional analog filter to the acquired waveforms, and designing the overall hybrid analog-digital system to facilitate target identification using task-based quantization methods. In particular, we exploit the fact that the target parameters can be recovered from a compressed representation of the received waveforms. We thus tune the acquisition system to recover this representation with minimal distortion, from which the targets can be extracted in digital, and characterize the achievable error in identifying the  targets. Our numerical results demonstrate that the proposed BiLiMO receiver operating with a bit budget of  one bit per sample achieves target recovery performance which approaches that of costly MIMO radars operating with unlimited resolution ADCs, while substantially outperforming MIMO receivers operating only in the digital domain under the same bit limitations.
\end{abstract}


%
\IEEEpeerreviewmaketitle

\vspace{-0.4cm}
\section{Introduction} 
\vspace{-0.1cm}
\IEEEPARstart{M}{ultiple}-input multiple-output (MIMO)  radar technology facilitates sensing with improved flexibility and performance compared to traditional phased-array radars \cite{ Li:MIMORadar-07, Jian2008MIMO}. These gains are achieved by employing multiple antenna elements at both the transmitter and receiver, and radiating a set of mutually orthogonal waveforms. 
While the theoretical gains of MIMO radar are well-established,  designing such a system gives rise to notable challenges in signal processing and hardware implementation due to its increased complexity.
These challenges impose a major drawback in emerging applications which are required to operate with limited power and cost-efficient hardware, including automotive radar \cite{Patole2017automotive}, unmanned aerial vehicle radar \cite{Rankin2015UAV}, and radar imaging for urban sensing \cite{Amin2008}. Consequently, there is a growing need to design MIMO radars in a cost-efficient manner, allowing the resulting system to comply with constraints on its power consumption, physical size and shape, and bandwidth.

A major source for the increased cost and power consumption of MIMO radar systems stems from the need to acquire and process multiple  signals while operating at large frequency bands.  Specifically, MIMO radars utilize a set of \acp{adc} at the receiver in order to convert the received waveforms into a finite-bit representation, such that they can be digitally processed.
The cost and energy consumption of an ADC grows rapidly with the sampling rate and the number of bits used for digital representation \cite{walden1999analog}. 
Therefore, when the number of antennas and the signal bandwidth is large, the cost and power consumption of these \acp{adc}  become prohibitive.
Furthermore, such acquisition generates massive data sets for representing the waveforms, whose processing and storage may induce a notable burden on the radar receiver.

The leading approach in the literature to facilitate MIMO radar with  low-rate \acp{adc} is to utilize \ac{cs} \cite{eldar2012compressed} in order to break the dependency of the sampling rate on the signal bandwidth.
Under this framework, a variety of sub-Nyquist sampling receivers \cite{Tropp:RD-10 ,Mishali:MWC-10,Xi:QuadCS-14}, as well as sub-Nyquist signal processing methods \cite{Baransky:subNyquistradar-14,Eldar:DoppFoc-14,Liu:QuadCS-15}, have been developed for radar applications; see  survey in \cite{Cohen2018-SubNyquistSPM}.
In particular, sub-Nyquist MIMO radar systems proposed in \cite{Rossi:SpatialCSMIMO-14,Cohen2018-TSP,Mishra2020TAES} utilize \ac{cs} tools to reduce the sampling rate and number of antennas in MIMO radar systems without compromising its sensing performance.  
These works mainly focus on reducing the sampling rate and ignore the quantization aspect of analog-to-digital conversion, assuming high-resolution quantizers, which tend to be costly and power hungry. 

Recent years have witnessed growing interest in signal processing systems operating with bit-limited quantizers. A common strategy to study signal processing with quantization constraints is to acquire the analog signals using low-resolution quantizers, and to compensate for the distortion induced in quantization via digital processing. Such digital processing strategies have been proposed in a multitude of different applications, including MIMO communications \cite{Mollen2016wideband}, channel estimation \cite{Wan2020LMMSE}, direction of arrival estimation \cite{Yu:DOA-SPL-2016,Liu2017onebit}, and spectrum sensing \cite{Fu:Asilomar-2017, Ren:MM-RELAX-TSP2019}.
In the context of radar systems, the works \cite{Xi2018-SAM,Li:Asilomar-2016,Zahabi:ICASSP-2017,Ameri2019onebitradar} and \cite{Xi2020gridless,xi2020joint} modified the processing to account for low-resolution quantized observations in pulse-Doppler radar and MIMO radar, respectively.
These works assume fixed one-bit quantizers which are ignorant of the system task, possibly with the addition of some dedicated time-varying reference signal to capture amplitude information \cite{Li:Asilomar-2016,Zahabi:ICASSP-2017,Wang2018onebit,Ameri2019onebitradar,Xi2020gridless}. The notable distortion induced in low rate task-ignorant quantization may severely affect the overall system performance.  

An alternative strategy to facilitate signal processing with quantization constraints is to account for the task for which the signal is acquired in quantization. Such {\em task-based quantization} schemes \cite{Shlezinger-TSP2019,shlezinger2018asymptotic,Shlezinger-2020taskbased,salamatian2019task,Wang-2019DMA,shlezinger2019deep,shlezinger2020learning}  exploit the fact that analog signals are commonly acquired not to be reconstructed, but in order to extract some lower-dimensional information from them. By doing so, task-based quantizers are typically capable of achieving improved performance in carrying out their associated tasks under bit constraints compared to purely digital processing \cite{Shlezinger-TSP2019,shlezinger2018asymptotic,Shlezinger-2020taskbased,salamatian2019task}. This is achieved  by designing the overall acquisition system in light of the task, incorporating analog pre-quantization processing, which results in a hybrid analog and digital (HAD) architecture. Such  HAD architectures are commonly utilized in MIMO communications \cite{mendez2016hybrid,ioushua2019family} and MIMO radar \cite{Chahrour2018hybrid} as a method to reduce the number of costly RF chains. The fact that
HAD systems are commonly used in MIMO radar, and that such systems acquire their received waveforms for a specific task, i.e., extracting the parameters of the targets, motivates the design of bit-limited MIMO radar receivers as task-based quantization systems, which is the purpose of the current work.    

Here we propose the \ac{bilimo} receiver, which is designed to accurately recover its targets while operating with bit constraints using task-based quantization methods.   \ac{bilimo}  follows a HAD architecture operating with conventional scalar uniform \acp{adc} and analog filters, in which both the analog and the digital components of the system are jointly designed to facilitate target recovery under quantization constraints. In particular, our design builds upon the insight that target identification can be represented as a sparse recovery task. Therefore,  \ac{bilimo}  is designed to yield a compressed representation from which the targets can be recovered, such that the effect of the distortion induced in quantization on the ability to identify the targets is mitigated.

We present two designs of the \ac{bilimo} receiver: The first considers MIMO radar systems with monotone waveforms. For such setups we characterize the acquisition system which recovers the desired sufficient compressed representation in a manner which minimizes the mean-squared error (MSE) between the recovered compressed representation and optimal linear estimate of it from unquantized data. Then, we derive the \ac{bilimo} receiver for multitone waveforms. For this case, our resulting receiver can be shown to minimize the MSE under additional assumptions, which hold when the echos observed at different frequency bins are uncorrelated. As the \ac{bilimo} receiver detects the targets from its recovered compressed representation via \ac{cs} methods, we characterize the stability in identifying the targets using $\ell_1$ sparse recovery methods. Our numerical evaluations   demonstrate that the target parameters estimation accuracy of the proposed \ac{bilimo} receiver with a tight bit budget equivalent to one bit per sample approaches that of MIMO receivers operating with infinite resolution quantizers, and that it notably outperforms digital-only receivers operating under the same bit budget. 

The rest of the paper is organized as follows: 
Section~\ref{sec:System} introduces the MIMO radar model with HAD receivers. 
The \ac{bilimo} receiver architecture is described in Section~\ref{sec:Model}.
In Section~\ref{sec:Design}, the \ac{bilimo} receiver  design via task-based quantization is studied, and the target recovery performance is analyzed.
Sections \ref{sec:Sims} and  \ref{sec:conclusion} provide numerical simulations and concluding remarks, respectively.
Detailed proofs  are delegated to the appendix.

Throughout the paper,  
we use lower-case (upper-case) bold characters to denote vectors (matrices).
The $i$th element of a vector $\mathbf{x}$ is written as $(\mathbf{x})_i$.
Similarly, the $(i,j)$th element of a matrix $\mathbf{X}$ is $(\mathbf{X})_{i,j}$.
We use $\mathbf{I}_N$ for the $N \times N$ identity matrix.
$\mathbb{R}$ and $\mathbb{C}$ denote the sets of real and complex numbers.
$(\cdot)^T$, $(\cdot)^H$, $\textrm{Tr}(\cdot)$, and $\sign(\cdot)$ denote the matrix transposition, Hermitian transposition, trace, and sign operator, respectively. Finally,
$\lceil\cdot\rceil$ and $\lfloor\cdot\rfloor$ denote the ceiling and the floor functions, $a^{+} \triangleq \max(a,0)$, and $\textrm{vec}(\cdot)$ is the vectorization operator.

\vspace{-0.2cm}
\section{HAD MIMO Radar Model}
\label{sec:System}
\vspace{-0.1cm}
In this section we present the system model of MIMO radar with HAD receivers. We begin by formulating the transmitted and received waveforms in Subsection \ref{subsec:SystemSignal}. Then, we detail the problem of target recovery using a bit-constrained HAD receiver in Subsection \ref{subsec:SystemHAD}, based on which we formulate the \ac{bilimo} receiver architecture and its corresponding design problem formulation in the following section.

\vspace{-0.2cm}
\subsection{MIMO Radar Signal Model}
\label{subsec:SystemSignal}
\vspace{-0.1cm}
We consider a colocated MIMO radar consisting of two linear antenna arrays with $N$ receive antennas and $M$ transmit antennas.
The locations of the $N$ receive antennas and $M$ transmit antennas are denoted by $\zeta_0\lambda, \cdots, \zeta_{N-1}\lambda$ and $\xi_0\lambda, \cdots, \xi_{M-1}\lambda$, respectively.
Here, $\lambda$ is the wavelength of the carrier signal.
Without loss of generality, we assume that $\zeta_0=\xi_0=0$.
The MIMO radar uses two uniform linear arrays (ULAs)  as the receive antennas and transmit antennas, located at $\zeta_n=n/2$ and $\xi_m=Nm/2$ for $0\leq n\leq N-1$ and $0\leq m\leq M-1$.
As such, the resulting $MN$ channels can generate a virtual ULA array with length $MN\lambda/2$ \cite{Li:MIMORadar-07}.

Each transmit antenna sends out $Q$ pulses, such that the $m$th transmitted signal is given by
\begin{equation}
	s_m(t) = \sum_{q=0}^{Q-1}h_m(t-qT_0)e^{j2\pi f_ct}, \quad 0\leq t\leq QT_0,
\end{equation}
where $h_m(t)$, $0\leq m\leq M-1$, are narrowband pulses with bandwidth $B_h$, modulated with carrier frequency $f_c$, and 
$T_0$ denotes the pulse repetition interval (PRI).
For simplicity, we only consider one PRI, i.e., $Q=1$.
However, our analysis can be generalized to the case of multiple pulses, namely, $Q>1$.

MIMO radar architectures commonly utilize orthogonal waveforms for radar probing \cite{Li:MIMORadar-07}.
Here, we consider orthogonality achieved using frequency division multiple access (FDMA) signaling. 
In FDMA, the transmitted baseband waveform $h_m(t)$ can be expressed as
	$h_m(t)=h_0(t)e^{j2\pi f_mt}$,
where $h_0(t)$ is the lowpass waveform with spectral support $[-\frac{B_h}{2},\frac{B_h}{2}]$, each $f_m$ is chosen in $[-\frac{MB_h}{2},\frac{MB_h}{2}]$ so that the intervals $[f_m-\frac{B_h}{2},f_m+\frac{B_h}{2}]$ do not overlap. 
In such setups, different waveforms lie in distinct spectral bands.

The targets are represented as non-fluctuating point reflectors in the far field, and we let $K$ be the number of targets.
Each target is characterized by the following parameters: its reflection coefficient $\tilde{\alpha}_k$, its distance from the array origin $R_k$, and the azimuth angle relative to the array $\theta_k$.
We assume the targets lie in the radar unambiguous time-frequency region.

\begin{figure}
    \centering
    \includefig{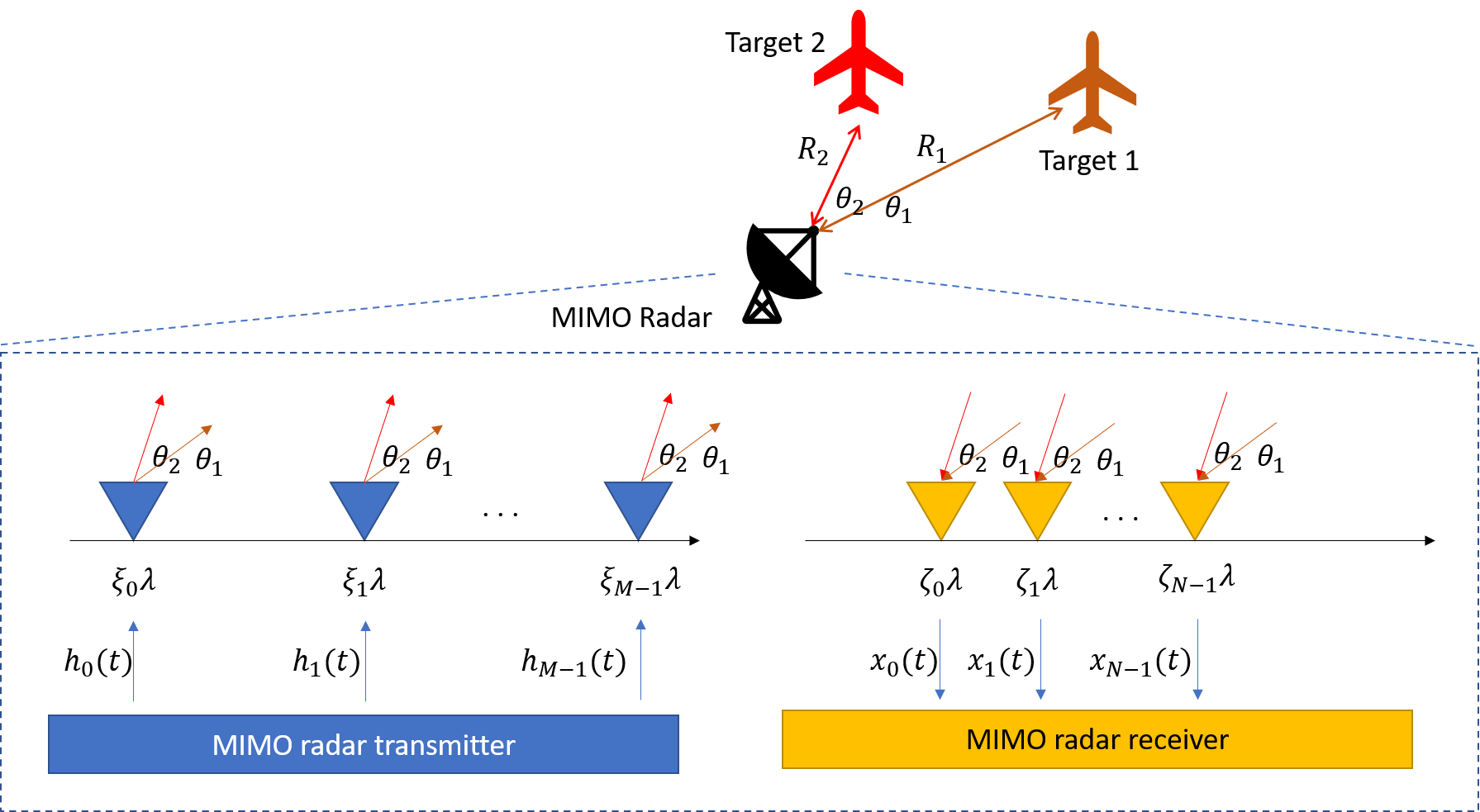}
    \vspace{-0.2cm}
    \caption{Baseband operation of a MIMO radar system with $K=2$ targets.}
    \label{fig:RadarSys}
\end{figure}

Let $\tau_k=\frac{2R_k}{c}$ and $\vartheta_k=\sin(\theta_k)$ be the delay and azimuth sine of the $k$th target, respectively.
The received signal $\tilde{x}_n(t)$ at the $n$th antenna in one PRI can then be written as:
\ifFullVersion
\begin{equation*}
   \tilde{x}_n(t) \!= \!\sum_{m=0}^{M-1}\sum_{k=1}^K\tilde{\alpha}_ke^{j2\pi f_c(\xi_m\!+\! \zeta_n)\vartheta_k}h_m(t\!-\!\tau_{k})e^{j2\pi f_c(t\!-\!\tau_{k})}\!+\!\tilde{w}_n(t), 
\end{equation*}
\else
$\tilde{x}_n(t) \!= \!\sum_{m=0}^{M-1}\sum_{k=1}^K\tilde{\alpha}_ke^{j2\pi f_c(\xi_m\!+\! \zeta_n)\vartheta_k}h_m(t\!-\!\tau_{k})e^{j2\pi f_c(t\!-\!\tau_{k})}\!+\!\tilde{w}_n(t)$, 
\fi
where $\tilde{w}_n(t)$ is the interference plus noise signal.
By defining $x_n(t)=\tilde{x}_n(t)e^{-j2\pi f_ct}$ as the baseband component, we have
\begin{equation}
	x_n(t)\triangleq\sum_{m=0}^{M-1}\sum_{k=1}^K\alpha_ke^{j2\pi(\xi_m+\zeta_n)\vartheta_k}h_m(t-\tau_{k})+w_n(t),\label{eqn:Xbase}
\end{equation}
with $\alpha_k=\tilde{\alpha}_ke^{-j2\pi f_c\tau_k}$ and $w_n(t) = \tilde{w}_n(t)e^{-j2\pi f_ct}$. In Fig.~\ref{fig:RadarSys} we illustrate our MIMO radar model for $K=2$.

\vspace{-0.2cm}
\subsection{HAD Radar Receiver}
\label{subsec:SystemHAD}
\vspace{-0.1cm}
The goal of MIMO radar receiver processing is to identify the targets based on the received echos. In particular, the receiver is required to resolve the $K$ delay-azimuth pairs $\{\tau_k, \vartheta_k\}_{k=1}^K$ from the received signals $\{x_n(t)\}_{n=0}^{N-1}$.
In classic MIMO radar, after demodulation, the baseband components of the received signals $\{x_n(t)\}_{n=0}^{N-1}$ are converted from analog signals to digital representations using \acp{adc} which sample above the Nyquist rate and utilize high-resolution quantizers. The outputs of the \acp{adc} are then processed in the digital domain 
in order to estimate the delays and azimuth sines of the $K$ targets from the received signals \cite{Jian2008MIMO}.
The usage of high-resolution \acp{adc}, which assign a relatively large number of bits to represent each sample, induces minimal distortion \cite{gray1998quantization}, and thus the effect of quantization on radar signal processing is usually ignored. 
Nonetheless, the fact that the power consumption of \ac{adc} devices  grows exponentially with the number of bits assigned to each sample \cite{walden1999analog}, dramatically affects the power and cost of MIMO radar systems operating at high frequencies with a large number of receive antennas. 

\begin{figure}
    \centering
    \includefig{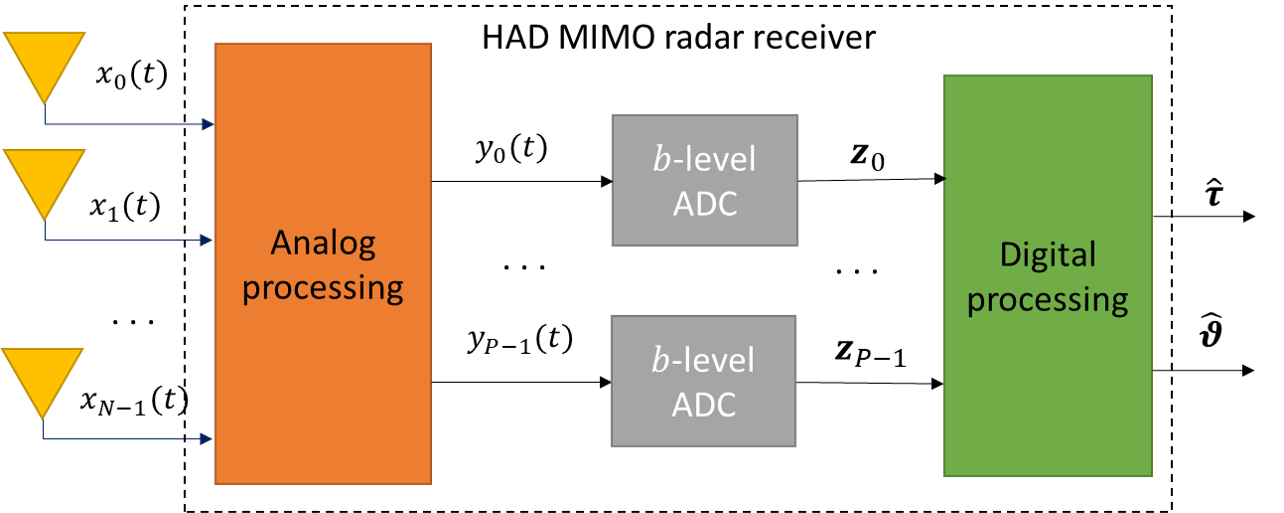}
    \caption{HAD radar receiver block diagram.}
    \label{fig:HAD_Rx1}
\end{figure}

Among the leading  design approaches to reduce the cost and power usage of such MIMO radar systems are $(i)$ utilize low-resolution ADCs; and $(ii)$ reduce the number of RF chains and ADCs by operating in a HAD manner. The usage of low-resolution ADCs implies that each ADC can output up to $b$ different levels, e.g., $b=4$ for two-bit ADCs. HAD architectures introduce pre-acquisition analog processing, combining the $N$ analog signals $\{x_n(t)\}_{n=0}^{N-1}$ into $P$ outputs $\{y_p(t)\}_{p=0}^{P-1}$, which are then acquired by the ADCs.  Setting $P < N$ implies that HAD systems reduce the number of costly RF chains and ADCs compared to conventional MIMO radar receivers. An illustration of such a HAD  receiver is depicted in Fig.~\ref{fig:HAD_Rx1}.

 Analog combining prior to analog-to-digital conversion is commonly studied in the MIMO communication literature \cite{ioushua2019family,mendez2016hybrid}, typically as a mean to reduce the number of costly RF chains.  HAD MIMO receivers were also shown to facilitate operation with low resolution quantizers for communication tasks \cite{shlezinger2018asymptotic,Wang-2019DMA}. This is achieved using task-based quantization methods \cite{Shlezinger-TSP2019,shlezinger2018asymptotic,Shlezinger-2020taskbased,salamatian2019task,Wang-2019DMA}, which tune the acquisition mapping in light of the overall system task, allowing to accurately recover the desired information under limited bit budgets. This motivates the design of   HAD receivers for the task of recovering the target parameters as a form of task-based quantization.
 
In order to design such HAD radar receivers, one must first introduce some constraints on the feasible mappings of the components of the system in Fig. \ref{fig:HAD_Rx1}. The motivation for imposing such constraints is two-fold: First, they enforce the resulting system to correspond to architectures which are feasible in terms of hardware. For instance, while in principle the analog processing in Fig.~\ref{fig:HAD_Rx1} can be any mapping, in practice it is likely to be implemented using analog hardware based on filters and multiplexers. The second motivation for introducing these constraints is to obtain an analytically tractable design problem, which is very challenging due to the complex relationship between the target parameters $\{\tau_k, \vartheta_k\}_{k=1}^K$ and the received signals $\{x_n(t)\}_{n=0}^{N-1}$. In the following section we introduce the considered constrained HAD radar receiver architecture, which we refer to as \ac{bilimo}. We then tackle the challenge in designing the receiver to recover the targets by formulating a relaxed problem, as shown in the sequel.


\vspace{-0.2cm}
\section{\ac{bilimo}  Receiver Architecture}
\label{sec:Model}
\vspace{-0.1cm}
In this section, we introduce the proposed \ac{bilimo}  receiver, which is designed to operate with bit budget constraints. Typical systems involving the acquisition of analog signals and their processing in the digital domain attempt to achieve an accurate digital representation of the acquired signals, and are thus prone to notable distortion when utilizing low-resolution ADCs.  
Here, we design \ac{bilimo}  as a HAD MIMO receiver, whose components are jointly designed to facilitate target recovery. The fact that the received signals in HAD systems are processed in analog prior to being converted to digital, facilitates extracting the desired information from them, as they can be combined into a lower dimensional representation from which the desired parameters are still recoverable. 

\begin{figure*}
\centering
\includegraphics[width=0.9\linewidth]{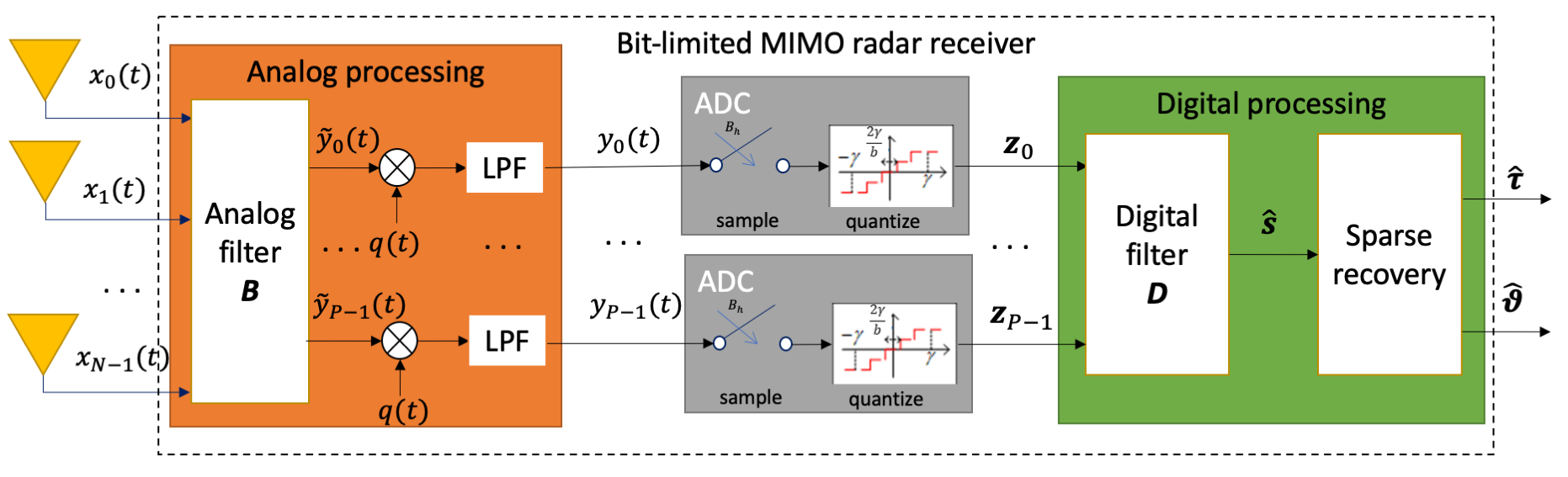}
\vspace{-0.5cm}
\caption{\ac{bilimo}  receiver illustration.}
\label{weightfunction}
\vspace{-0.4cm}
\end{figure*}

In particular, \ac{bilimo} is a HAD MIMO receiver, as illustrated in Fig.~\ref{fig:HAD_Rx1}, designed in light of the additional constraints: 
The analog processing is implemented using a set of combining filters, mixers, and low-pass filters, as detailed in Subsection~\ref{subsec:APP}; analog-to-digital conversion is carried out using identical uniform ADCs, as discussed in Subsection~\ref{subsec:ADC}. Due to the complex relationship between the target parameters and the observations, we formulate a relaxed problem of digitally filtering the ADCs output to recover a lower-dimensional representation that preserves the semantic information with respect to the target parameters. The targets can then be recovered using further digital processing based on conventional sparse recovery mechanisms. This digital processing is detailed in Subsection~\ref{subsec:DP}, and the relaxed problem is formulated in Section~\ref{sec:Design}. The resulting architecture of the \ac{bilimo} receiver is illustrated in Fig.~\ref{weightfunction}.

\vspace{-0.2cm}
\subsection{Analog Pre-Processing}
\label{subsec:APP}
\vspace{-0.1cm}
The analog processing consists of three stages: analog combining, analog mixing, and filtering.
First, the $N$ received signals are combined to output $P$ ($P\leq MN$) channels.
Then each of the $P$ channels are separately mixed with a mixing signal and low-pass filtered before being fed to the ADCs. The motivation for using this  architecture stems from the fact that it equivalently implements channel separation, which is typically the first step in MIMO radar processing required to achieve its desired virtual array capabilities, followed by a controllable analog combiner. However, while the direct implementation of such analog hardware requires $M N$ bandpass filters (for channel separation) followed by $MNP$ controllable filters, each applied to a different bandpass component, the architecture illustrated as {\em Analog processing} in Fig. \ref{weightfunction} can be shown to implement the same mapping while utilizing merely $NP$ filters applied directly to the full-band received signals, followed by $P$ identical low-pass filters.

Let $b_{p,n}(t)$ be the $(p,n)$th analog filter.
The $p$th output of the analog combining can be expressed as  
\ifFullVersion
\begin{align}
     &\tilde{y}_p(t) =\sum_{n=0}^{N-1} b_{p,n}(t)\ast x_n(t)\notag \\
     &=\sum_{n=0}^{N-1} \sum_{m=0}^{M-1} \sum_{k=1}^K \alpha_k e^{j2\pi(\xi_m+\zeta_n)\vartheta_k} [b_{p,n}(t)\ast h_m(t-\tau_{k})].
\end{align}
\else
$\tilde{y}_p(t) =\sum_{n=0}^{N-1} b_{p,n}(t)\ast x_n(t)$, resulting in $\tilde{y}_p(t) =\sum_{m=0}^{M-1} \sum_{k=1}^K \alpha_k e^{j2\pi(\xi_m+\zeta_n)\vartheta_k} [b_{p,n}(t)\ast h_m(t-\tau_{k})]$.
\fi 
Since $\tilde{y}_p(t)$ is limited to $t \in[0, T_0]$, it can be equivalently expressed by its Fourier series 
\begin{equation}\label{eqn_FS}
	\tilde{y}_p(t)=\sum_{i=\lceil -MT_0B_h/2\rceil}^{\lfloor MT_0B_h/2\rfloor} \tilde{c}_{p}[i]e^{j2\pi it/T_0}, \quad t\in[0,T_0],
\end{equation}
 where
\ifFullVersion
\begin{align} 
    &\tilde{c}_{p}[i]=\frac{1}{T_0}\int_0^{T_0} \tilde{y}_p(t)e^{-j2\pi it/T_0}dt \notag \\
    &\!=\!\!\sum_{n=0}^{N-1}\!\sum_{m=0}^{M-1}\hat{h}_m\!\left(\!\frac{2\pi i}{T_0}\!\right)\!\hat{b}_{p,n}\!\left(\!\frac{2\pi i}{T_0}\!\right)\!\sum_{k=1}^K\frac{\alpha_k}{T_0}e^{j2\pi\left((\xi_m\!+\!\zeta_n)\vartheta_k\!-\frac{ i \tau_k}{T_0}\right)}\notag, 
\end{align}
\else
$\tilde{c}_{p}[i]=\frac{1}{T_0}\int_0^{T_0} \tilde{y}_p(t)e^{-j2\pi it/T_0}dt$, which equals $\tilde{c}_{p}[i] \!=\!\!\sum_{n=0}^{N-1}\!\sum_{m=0}^{M-1}\hat{h}_m\!\left(\!\frac{2\pi i}{T_0}\!\right)\!\hat{b}_{p,n}\!\left(\!\frac{2\pi i}{T_0}\!\right)\!\sum_{k=1}^K\!\frac{\alpha_k}{T_0}e^{j2\pi\left((\xi_m\!+\!\zeta_n)\vartheta_k\!-\frac{ i \tau_k}{T_0}\right)}$,
\fi
where  $\hat{b}_{p,n}(\omega)$ and $\hat{h}_m(\omega)$ denote the Fourier transform of $b_{p,n}(t)$ and $h_m(t)$, respectively.

Each of the $P$ output channels is mixed with the signal $q(t)=\sum_{m=0}^{M-1}e^{-j2\pi f_m t}$ and filtered by a lowpass filter with passband $[-\pi{B_h},\pi{B_h}]$.
The motivation of using such mixing signals is to combine the spectrum of the $M$ transmitted signals, such that a portion of energy from
each band appears in baseband. Its combination with low-pass filtering results in equivalent outputs to those which would have been produced by first applying channel separation based on $MN$ matched filters, as shown in the sequel, without having to implement these channel separation filters in analog hardware.
This structure is similar to that used in Xampling \cite{Cohen2018-TSP}, which has been shown to be conveniently implemented in hardware. 
Using \eqref{eqn_FS}, the   output of the $p$th channel is
\begin{equation}\label{eqn_LPF}
        y_p(t)
        =\sum_{m=0}^{M-1}\sum_{i=\lceil -T_0B_h/2\rceil}^{\lfloor T_0B_h/2\rfloor}  \tilde{c}_{p}[i+f_mT_0]e^{j2\pi it/T_0}.
\end{equation}

Let us define
\begin{equation}\label{eqn_b_pnm}
    \hat{b}_{p,n}^{m}(\omega)\triangleq \begin{cases}
              \frac{1}{T_0}\hat{h}_0(\!\omega)\hat{b}_{p,n}(\!\omega\!+\!2\pi\! f_m)  &\!\omega\!\in\![\!-\!\pi \!B_h,\!\pi\! B_h\!]\\
               0 & \text{else}.\\
               \end{cases}
\end{equation}
Then, for $\lceil -T_0B_h/2\rceil \leq i \leq \lfloor T_0B_h/2\rfloor$,
\ifFullVersion
$\tilde{c}_{p}[i+f_mT_0]$ can be expressed as
\begin{equation}
    \tilde{c}_{p}[i+f_mT_0] = \sum_{n=0}^{N-1}c_{m,n}[i]b_{p,mN+n}[i],
\end{equation}
\else 
it holds that $ \tilde{c}_{p}[i+f_mT_0] = \sum_{n=0}^{N-1}c_{m,n}[i]b_{p,mN+n}[i]$, 
\fi
where 
\begin{equation}\label{eqn_cmni}
    c_{m,n}[i]\triangleq\sum_{k=1}^K\alpha_k e^{j2\pi\left((\xi_m\!+\!\zeta_n)\vartheta_k\!-\frac{ i \tau_k}{T_0}-f_m\tau_k\right)},
\end{equation}
and $b_{p,mN+n}[i]\triangleq \hat{b}_{p,n}^{m}\left(\frac{2\pi i}{T_0}\right)$. 
Define $L\triangleq B_hT_0$, assumed to be an odd integer in the following discussion for convenience.
Then the output in \eqref{eqn_LPF} can be written as
\begin{equation}\label{eqn_yp2}
    y_p(t)=\sum_{i=-\frac{L-1}{2}}^{\frac{L-1}{2}}\sum_{n=0}^{N-1}\sum_{m=0}^{M-1}b_{p,mN+n}[i]c_{m,n}[i]e^{j2\pi it/T_0},
\end{equation}
which is equivalent to the outputs achieved by applying channel separation followed by an analog combiner comprised of $MNP$ individual filters with the frequency responses in \eqref{eqn_b_pnm}. 

We can now write the output of the analog processing component in multivariate form by defining  $\mathbf{y}(t)\triangleq[y_{0}(t),\cdots,y_{P-1}(t)]^T$, and obtaining from \eqref{eqn_yp2} that
\ifFullVersion
\begin{equation}
\label{eqn_yp3}
	\mathbf{y}(t) = \sum_{i=-\frac{L-1}{2}}^{\frac{L-1}{2}}e^{j2\pi it/T_0} \mathbf{B}_i\mathbf{c}_i,
\end{equation}
\else
$	\mathbf{y}(t) = \sum_{i=-\frac{L-1}{2}}^{\frac{L-1}{2}}e^{j2\pi it/T_0} \mathbf{B}_i\mathbf{c}_i$, 
\fi 
where $\mathbf{B}_i\triangleq[b_{p,\tilde{n}}[i]]\in\mathbb{C}^{P\times MN}$ with $0\leq \tilde{n} < MN$, and $\mathbf{c}_i\triangleq[\mathbf{c}_0^T[i],\cdots,\mathbf{c}_{M-1}^T[i]]^T$ with $\mathbf{c}_m[i]=[c_{m,0}[i],\cdots,c_{m,N-1}[i]]^T$ for each $0\leq m < M$. 
By using the above analog processing, rather than classic matched filtering, the number of channels to be processed is reduced from $MN$ to $P$, facilitating the sampling and quantization operations.
The  design  of  the  analog  processing is equivalent to designing the $NP$ analog filters $b_{p,n}(t)$, which can be constructed from the desired value of $\mathbf{B}_i$. 
Specifically, the frequency response of $b_{p,n}(t)$ becomes $\hat{b}_{p,n}(2\pi(\frac{i}{T_0}+f_m))=T_0b_{p,mN+n}[i]\hat{h}_0^\ast(\frac{2\pi i}{T_0})/|\hat{h}_0(\frac{2\pi i}{T_0})|^2$ for $-\frac{L-1}{2}\leq i\leq \frac{L-1}{2}$ by \eqref{eqn_b_pnm}. 
After analog processing, the $P$ signals represented by $\mathbf{y}(t) $ are forwarded to the \acp{adc} as detailed in the following.

\vspace{-0.2cm}
\subsection{Analog-to-Digital Conversion}
\label{subsec:ADC}
\vspace{-0.1cm}
The output of the analog processing  $\mathbf{y}(t) $ is converted into a set of digital streams via sampling and quantization. This conversion is carried out using $P$ identical pairs of \acp{adc} which independently discretize the real and imaginary parts of each analog input signal.
We focus here on the low-resolution quantization aspect of analog-to-digital conversion, assuming that Nyquist sampling is applied, and leave the analysis and joint design of such systems with sub-Nyquist sampling and low resolution quantization for future work.

To formulate the \ac{adc} operation, we let $\mathbf{y}_i\triangleq \mathbf{y}(\frac{i}{B_h})$, $i=0,\cdots,L-1$, be the Nyquist samples of $\mathbf{y}(t)$.
Define the $PL\times 1$ samples vector $	\mathbf{y}\triangleq[\mathbf{y}_0^T,\cdots,\mathbf{y}_{L-1}^T]^T$, and the $MNL\times 1$ Fourier coefficients vector as $\mathbf{c}\triangleq[\mathbf{c}_{-(L-1)/{2}}^T, \cdots,\mathbf{c}_{(L-1)/{2}}^T]$.  
Using these definitions, the samples of the outputs of the analog combining filters  can be expressed as
\begin{equation}
	\mathbf{y}=\bar{\mathbf{F}}\bar{\mathbf{B}}\mathbf{c},
	\label{eqn:SigInput}
\end{equation}
where $\bar{\mathbf{B}}\triangleq\blkdiag(\mathbf{B}_{-(L-1)/2},\cdots,\mathbf{B}_{(L-1)/2})\in\mathbb{C}^{PL\times MNL}$, 
$\bar{\mathbf{F}}\triangleq\mathbf{F}_{L}^H\otimes \mathbf{I}_{P}$, and $\mathbf{F}_{L}$ is the $L\times L$ \ac{dft} matrix.

Using a similar derivation, we represent the samples of the interference and noise signal as
\ifFullVersion
\begin{equation}
	\mathbf{n} = \bar{\mathbf{F}}\bar{\mathbf{B}}\mathbf{w},
	\label{eqn:NoiseInput}
\end{equation}
\else
$\mathbf{n} = \bar{\mathbf{F}}\bar{\mathbf{B}}\mathbf{w}$, 
\fi
where $\mathbf{w}\in\mathbb{C}^{MNL}$ is the frequency-domain representation of the interference and noise signal observed at the $N$  antennas.

The sampled signal $\mathbf{y}$ is converted into a digital representation using uniform complex-valued quantizers with $b$ decision regions and support $\gamma>0$. The resulting quantization mapping is given by $\mathcal{Q}_C^{\gamma,b}(\cdot)=\mathcal{Q}^{\gamma,b}(\Re\{\cdot\})+j\mathcal{Q}^{\gamma,b}(\Im\{\cdot\})$, where $\mathcal{Q}^{\gamma,b}(\cdot)$ is the real-valued quantization operator applied element-wise to any real vector or matrix, and is given by
\begin{equation*}
	\mathcal{Q}^{\gamma,b}(x)=\begin{cases}
		-\gamma+\frac{2\gamma}{b}(l+\frac{1}{2}) &\begin{array}{l}
		      x-l\frac{2\gamma}{b}+\gamma\in[0,\frac{2\gamma}{b}], \\
		     l\in\{0,1,\cdots,b-1\},
		\end{array} \\
		\sign(x)(\gamma-\frac{\gamma}{b}) & |x|>\gamma.
	\end{cases}
\end{equation*}
The output of the \acp{adc} is the vector $\mathbf{z}\in\mathbb{C}^{PL}$, given by
\begin{equation}\label{eqn_model}
	\mathbf{z} = \mathcal{Q}_C^{\gamma,b}(\mathbf{y}+\mathbf{n})  
	=\mathcal{Q}_C^{\gamma,b}(\bar{\mathbf{F}}\bar{\mathbf{B}}\mathbf{c}+\bar{\mathbf{F}}\bar{\mathbf{B}}\mathbf{w}). 
\end{equation} 

In our design of the \ac{bilimo} receiver in Section \ref{sec:Design} we model the \acp{adc} as implementing non-subtractive dithered quantization \cite{gray1993dithered}. This model facilitates the design and analysis of bit-constrained HAD systems \cite{Shlezinger-TSP2019}, while being a faithful approximation of conventional uniform quantization operation under various statistical models \cite{widrow1996statistical}.
The number of bits used for representing $\mathbf{z}$ is thus $2P \lceil\log b \rceil$. Processing of $\mathbf{z}$ in the digital domain is detailed in the following subsection.

\vspace{-0.2cm}
\subsection{Digital Processing}
\label{subsec:DP}
\vspace{-0.1cm}
The digital representation $\mathbf{z}$ is used to recover the target parameters.
However, the relationship between $\mathbf{z}$ \eqref{eqn_model} and the target parameters $\{\tau_k, \vartheta_k\}_{k=1}^K$ is quite complex, making the joint design of the analog processing and digital mapping very difficult. Therefore, in the following we partition the digital processing into two stages, as illustrated in Fig.~\ref{weightfunction}. The first part is comprised of a digital filter, which is jointly designed with the analog combiner and ADC support based on a {\em relaxed problem} of recovering a vector $\mathbf{s}$ in the sense of minimal MSE. The relaxed task vector $\mathbf{s}$ is selected such that the target parameters can be recovered from it using conventional linear sparse recovery algorithms in the second part of the digital processing.  The \ac{bilimo} receiver thus builds upon the ability to recast  target identification  as a sparse recovery problem \cite{de2019compressed}. Therefore, to  formulate the problem of designing the \ac{bilimo} receiver, we first  rewrite our parameter estimation task as sparse recovery, after which we present the resulting digital processing structure, which is designed based on the relaxed objective detailed in Section~\ref{sec:Design}.

As in classic MIMO radar, we now assume the parameters $\tau_k$ and $\vartheta_k$ are located on the Nyquist grid, i.e., $\tau_k\in\{\frac{T_0l}{ML}\}_{l=0}^{ML-1}$ and $\vartheta_k\in\{-1+\frac{2l}{MN}\}_{l=0}^{MN-1}$.
It follows from \eqref{eqn_cmni} that the vector $\tilde{\mathbf{c}}_m\triangleq[(\mathbf{c}_m[-\frac{L-1}{2}])^T,\cdots,(\mathbf{c}_m[\frac{L-1}{2}])^T]\in\mathbb{C}^{NL}$ 
obeys the following sparse representation 
\ifFullVersion
\begin{align} 
    \tilde{\mathbf{c}}_m &= \vect(\mathbf{U}_m\TarMat\mathbf{V}_m^T)
    =(\mathbf{V}_m\otimes \mathbf{U}_m)\TarVec, \label{eqn:SparseRep1}
\end{align}
\else
$\tilde{\mathbf{c}}_m = \vect(\mathbf{U}_m\TarMat\mathbf{V}_m^T)    =(\mathbf{V}_m\otimes \mathbf{U}_m)\TarVec$,
\fi
where $\mathbf{U}_m\in\mathbb{C}^{N\times MN}$ with $(\mathbf{U}_m)_{n,l}=e^{j2\pi(\xi_m+\zeta_n)(-1+\frac{2l}{MN})}$, $\mathbf{V}_m\in\mathbb{C}^{L\times MN}$ with $(\mathbf{V}_m)_{i,l}=e^{-j2\pi( i/T_0+f_m)\frac{T_0l}{ML}}$, $\TarMat\in\mathbb{C}^{MN\times ML}$ is a sparse matrix that contains $K$ nonzero elements,
 and $\TarVec=\vect(\TarMat)\in\mathbb{C}^{M^2NL}$ is thus a $K$-sparse vector. {The sparsity pattern of $\TarVec$ encapsulates the values of the unknown delays and angles, i.e., if the $k$th non-zero element of $\TarVec$ is located at its index $(l_1-1)MN + l_2$ where $0\leq l_1 < ML$ and $0 \leq l_2 < MN$, then $\tau_k = \frac{T_0l_1}{ML}$ and $\vartheta_k=-1+\frac{2l_2}{MN}$}.

Stacking $\{\tilde{\mathbf{c}}_m\}_{m=1}^{M-1}$ into the $MNL\times 1$ vector $\tilde{\mathbf{c}}$, we obtain
\begin{equation}\label{eqn_c}
	\tilde{\mathbf{c}}=\mathbf{\Phi}\TarVec,
\end{equation}
where $\mathbf{\Phi} \in \mathbb{C}^{MNL\times M^2NL}$ is defined as
\ifFullVersion
\begin{equation}\label{eqn_Phi}
	\mathbf{\Phi}\triangleq[(\mathbf{V}_0^T\!\otimes\! \mathbf{U}^T_0)),(\mathbf{V}^T_1\!\otimes\! \mathbf{U}^T_1), \cdots,(\mathbf{V}^T_{M\!-\!1}\!\otimes\! \mathbf{U}^T_{M\!-\!1})]^T.
\end{equation}
\else
$	\mathbf{\Phi}\triangleq[(\mathbf{V}_0^T\!\otimes\! \mathbf{U}^T_0)),  \cdots,(\mathbf{V}^T_{M\!-\!1}\!\otimes\! \mathbf{U}^T_{M\!-\!1})]^T$.
\fi
Using the sparse representation \eqref{eqn_c}, the quantized $\mathbf{z}$ \eqref{eqn_model} becomes
\ifFullVersion
\begin{equation}\label{eqn_modelSparse} 
	\mathbf{z} = \mathcal{Q}_C^{\gamma,b}(\bar{\mathbf{F}}\bar{\mathbf{B}}(\mathbf{P}\mathbf{\Phi}\TarVec+\mathbf{w})),  
\end{equation}
\else
$	\mathbf{z} = \mathcal{Q}_C^{\gamma,b}(\bar{\mathbf{F}}\bar{\mathbf{B}}(\mathbf{P}\mathbf{\Phi}\TarVec+\mathbf{w}))$, 
\fi
where $\mathbf{P}\in\mathbb{R}^{MNL\times MNL}$ is a permutation matrix which aligns the elements between $\mathbf{c}$ and $\tilde{\mathbf{c}}$, i.e., $\mathbf{c}=\mathbf{P}\tilde{\mathbf{c}}$.

The MIMO radar task is thus equivalent to recovering the $K$-sparse vector $\TarVec$ from the quantized $\mathbf{z}$. Therefore, to facilitate the joint design of \ac{bilimo} as a HAD receiver, we set its digital processing to first recover a $J \times 1$ vector $\mathbf{s}$, with $J\leq PL$, which can be written as a linear compressed representation of $\TarVec$. This is achieved by first applying a digital filter  $\mathbf{D}\in \mathbb{C}^{J\times PL}$, which can be designed using task-based quantization methods, as we show in the following section, such that its output $\hat{\mathbf{s}} = \mathbf{D}\mathbf{z}$ is an accurate estimate of   $\mathbf{s}$. Then, the fact that $\hat{\mathbf{s}}$ can be written as a linear compressed representation of $\TarVec$ with some additive estimation error term is exploited in the subsequent digital processing, which resolves the targets from $\hat{\mathbf{s}}$ using conventional algorithms for recovering sparse vectors from noisy linear compressed measurements. 
The formulation of the relaxed problem and the resulting design of \ac{bilimo} are detailed in the following section. 




\color{black}

\vspace{-0.2cm}
\section{\ac{bilimo} Receiver Design}
\label{sec:Design}
\vspace{-0.1cm}
In this section we jointly design the HAD processing and quantizer mapping of the \ac{bilimo} receiver for target recovery under bit constraints. Our approach builds upon the task-based quantization framework proposed in \cite{Shlezinger-TSP2019}.
We  first present a relaxation of the target detection problem which represents the design of \ac{bilimo} as task-based quantization in Subsection \ref{subsec:problem}. 
Then, we show how this formulation allows designing the  \ac{bilimo} receiver in Subsections \ref{subsec:Design1}-\ref{subsec:Design2}. Finally, we derive bounds on the target recovery accuracy  and discuss the  design in Subsections~\ref{subsec:Analysis}-\ref{subsec:Discussion}, respectively.

\vspace{-0.2cm}
\subsection{Optimization Problem Formulation}
\label{subsec:problem}
\vspace{-0.1cm}
Task-based quantization is a framework for designing HAD acquisition systems operating under bit constraints. Such methods aim to facilitate the recovery of information embedded in the observed analog signals, rather than preserving sufficiency of the digital representation with respect to the signal itself \cite{Shlezinger-2020taskbased}. In particular, the work \cite{Shlezinger-TSP2019} jointly designed the components of an acquisition systems including analog pre-quantization filtering and digital linear processing of a similar structure as those in \ac{bilimo} for the task of recovering a linear function of the measurements. However, unlike the setup in \cite{Shlezinger-TSP2019}, the task of the \ac{bilimo} receiver is to recover the parameters of the targets, which do not obey a linear form. 
To encompass this challenge in utilizing task-based quantization tools for MIMO radar, we next present a relaxed problem formulation, which decomposes target identification into a linear recovery problem followed by a linear sparse recovery task. This relaxation allows the former to be treated using existing results in task-based quantization theory, and the latter be tackled using conventional \ac{cs} methods.

In the proposed \ac{bilimo} receiver, rather than recovering the $K$-sparse vector $\TarVec$ in \eqref{eqn_c} directly, a digital filter is first applied to estimate a compressed vector $\mathbf{s}\in\mathbb{C}^{J}$ ($J\leq PL$) from the quantized data $\mathbf{z}$. In particular, we set $\mathbf{s}=\mathbf{M}\mathbf{\Phi}\TarVec$, i.e., $\mathbf{s}$ is a linear compressed representation of the desired  $\TarVec$, where $\mathbf{M}\in\mathbb{C}^{J\times MNL}$ is a pre-defined compressive measurement matrix. 
Then, the $K$-sparse vector $\TarVec$ is recovered from the estimate of $\mathbf{s}$ by applying sparse recovery techniques.


Although we refer to $\mathbf{s}$ as the {\em task} vector when applying task-based quantization tools, the true task of the system is to recover $\TarVec$, and the estimation of $\mathbf{s}$ is an intermediate step in that aim. Therefore, the setting of $\mathbf{s}$ can be treated as part of the design procedure. Specifically, the $J$-dimensional vector $\mathbf{s}$ is related to the $MNL$-dimensional vector $\mathbf{\Phi}\TarVec$ via the compressive measurement matrix $\mathbf{M}$. As a result, $\mathbf{M}$ should be selected such that the desired $\TarVec$ is still recoverable, while allowing \ac{bilimo} to obtain an accurate estimate of $\mathbf{s}$  at the first stage of its digital processing. The additional dimenionality reduction induced by $\mathbf{M}$  can be translated into improved accuracy when jointly designing the HAD system including the digital filter $\mathbf{D}$ to estimate $\mathbf{s}$ via task-based quantization. In particular, the accuracy of task-based quantization typically improves when the task dimensionality is reduced, as the same number of bits can be utilized to recover less quantities via HAD processing \cite{Shlezinger-TSP2019}. Thus, our relaxed problem formulation considers the estimation of the further compressed   $\mathbf{s}$, from which the targets are still recoverable via sparse recovery, rather than $\mathbf{\Phi}\TarVec$. In the following we formulate our problem for a given $\mathbf{M}$, providing guidelines for its setting and numerically evaluating different selections in Subsection~\ref{subsec:Analysis} and Section~\ref{sec:Sims}, respectively.

The relaxed objective of the jointly designed hybrid acquisition system is therefore to minimize the MSE  between the compressed vector $\mathbf{s}$ and the digital filter output, given by
\vspace{-0.1cm}
\begin{equation}
\label{eqn:Shat}
    \hat{\mathbf{s}}=\mathbf{D}\mathcal{Q}_C^{\gamma,b}(\bar{\mathbf{F}}\bar{\mathbf{B}}(\mathbf{P}\mathbf{\Phi}\TarVec+\mathbf{w})).
\vspace{-0.1cm}
\end{equation}
The resulting analysis characterizes the corresponding HAD processing and low-bit quantizers which achieve this MSE.
In particular, we assume that the \ac{bilimo} receiver has knowledge of:
1) the statistical model of $\TarVec$ and $\mathbf{w}$; 
2) the compressive measurement matrix $\mathbf{M}$, which allows recovery of $\TarVec$ from $\mathbf{s}$.

Quantizers are typically designed to operate within their dynamic range, namely, that their input lies within the support $[-\gamma,\gamma]$, to avoid inducing additional distortion due to saturation \cite{gray1998quantization}. Our derivation is thus carried out assuming that non-overloaded quantizers, i.e., the magnitudes of the real and imaginary parts of $\mathbf{z}$ are not larger than  $\gamma$  with sufficiently large probability.
To guarantee this, we fix $\gamma$ to be some multiple $\eta$ of the maximal standard deviation of the inputs:
\vspace{-0.1cm}
\begin{equation}\label{eqn_gamma}
	\gamma^2=\eta^2\max_{l=1,2,\cdots,P}\mathbb{E}\big\{\big|(\mathbf{y}+\mathbf{n})_l\big|^2\big\}.
\vspace{-0.1cm}
\end{equation}
For instance, for proper-complex Gaussian inputs, setting $\eta \geq \sqrt{2}$ yields overload probability smaller than $6\%$
\cite{shlezinger2018asymptotic}. For arbitrary inputs, one can set $\eta$ to obtain a desired  overload probability bound   via Chebyshev's inequality \cite[Pg. 64]{cover2012elements}. 

Following the framework of task-based quantization in \cite{Shlezinger-TSP2019}, we aim  to jointly design the analog combining matrix $\bar{\mathbf{B}}$, the digital processing matrix $\mathbf{D}$, and the support of the quantizer $\gamma$ via \eqref{eqn_gamma}, such that $ \hat{\mathbf{s}}$ approaches the linear minimal MSE (LMMSE) estimator of $\mathbf{s}$ from $\mathbf{P}\mathbf{\Phi}\TarVec+\mathbf{w}$, denoted $\tilde{\mathbf{s}}$. The motivation for this formulation stems from the fact that the output of the digital filter can be treated as an estimate of $\mathbf{s}$ from $\mathbf{P}\mathbf{\Phi}\TarVec+\mathbf{w}$ by \eqref{eqn:Shat}.  
Let $\LMMSEMat$ be the LMMSE transformation, i.e., $\tilde{\mathbf{s}}=\LMMSEMat(\mathbf{P}\mathbf{\Phi}\TarVec+\mathbf{w})$, and let $\mathbf{R}_{c}$ and $\mathbf{R}_{w}$ be the covariance matrices of $\mathbf{c}=\mathbf{P}\mathbf{\Phi}\TarVec$ and $\mathbf{w}$, respectively.
As $\tilde{\mathbf{s}}$ is the LMMSE estimate of $\mathbf{s}=\mathbf{M}\mathbf{\Phi}\TarVec$, it holds that $\LMMSEMat$ is given as
\vspace{-0.1cm}
\begin{equation}
	\LMMSEMat=\mathbf{M}\mathbf{P}^T\mathbf{R}_{c}\CovMat^{-1},
\vspace{-0.1cm}
\end{equation}
where $\CovMat=\mathbf{R}_{c} +\mathbf{R}_{w}$. Accordingly, the   LMMSE   $\LMMSE\triangleq\mathbb{E}\{\|\tilde{\mathbf{s}}\!-\!\mathbf{s}\|^2\}$ is  
\ifFullVersion
\begin{equation} 
   \LMMSE=\Tr\left[\mathbf{M}\mathbf{P}^T\mathbf{R}_{c}\mathbf{P}\mathbf{M}^H\!-\!\mathbf{M}\mathbf{P}^T\mathbf{R}_{c}\CovMat^{-1}\mathbf{R}_{c}\mathbf{P}\mathbf{M}^H\right].
   \label{eqn:LMMSE_Err}
\end{equation}
\else
$\LMMSE=\Tr\big[\mathbf{M}\mathbf{P}^T\mathbf{R}_{c}\mathbf{P}\mathbf{M}^H\!-\!\mathbf{M}\mathbf{P}^T\mathbf{R}_{c}\CovMat^{-1}\mathbf{R}_{c}\mathbf{P}\mathbf{M}^H\big]$.
\fi

To summarize, our goal is to optimize the components of the \ac{bilimo} receiver to minimize the excess MSE (EMSE):
\begin{equation}\label{prob_minMSE}
    \min_{\bar{\mathbf{B}}, \mathbf{D}, \gamma} \mathbb{E}\{\|\tilde{\mathbf{s}}-\hat{\mathbf{s}}\|^2\}.
\end{equation}
Our derivation of the  jointly designed \ac{bilimo} receiver is presented in the sequel, where we first focus on the special case of monotone waveforms, i.e., $L=1$, after which we show how these results extend to multitone waveforms with $L>1$.



\vspace{-0.2cm}
\subsection{\ac{bilimo} Receiver Design for Monotone Waveforms}
\label{subsec:Design1}
\vspace{-0.1cm}
We now characterize the \ac{bilimo} receiver design based on the objective \eqref{prob_minMSE}. In particular, we derive the analog combining matrix, digital processing matrix, and the support $\gamma$, for the scenario in which $L=1$, which means that monotone waveforms are transmitted.
In this case, the matrix $\bar{\mathbf{F}}$ in \eqref{eqn:SigInput}  is the identity matrix. As a result, the signal samples and noise samples can now be written as $\mathbf{y}=\mathbf{B}\mathbf{P}\mathbf{\Phi}\TarVec$ and $\mathbf{n}=\mathbf{B}\mathbf{w}$, respectively, where $\mathbf{B}=\mathbf{B}_0$ is the $P\times MN$ analog combining matrix.
Thus, the quantized output 
\ifFullVersion
$\mathbf{z}$ is given by
\begin{equation}\label{eqn_model1}
	\mathbf{z} = \mathcal{Q}_C^{\gamma,b}(\mathbf{B}(\mathbf{P}\mathbf{\Phi}\TarVec+\mathbf{w})).
\end{equation} 
\else
is $\mathbf{z} = \mathcal{Q}_C^{\gamma,b}(\mathbf{B}(\mathbf{P}\mathbf{\Phi}\TarVec+\mathbf{w}))$.
\fi

We begin by characterizing the digital processing matrix which minimizes the MSE for a fixed analog combining matrix $\mathbf{B}$.
By applying  \cite[Lemma 1]{Shlezinger-TSP2019}, we obtain the follow result:
\begin{lem}\label{lemma1} 
For any analog combining matrix $\mathbf{B}$, the digital processing matrix  which minimizes the MSE is given by
 \begin{equation}
	\mathbf{D}^\circ(\mathbf{B}) = \mathbf{M}\mathbf{P}^T\mathbf{R}_{c}\mathbf{B}^H\left(\mathbf{B}\CovMat\mathbf{B}^H+\frac{4\gamma^2}{3b^2}\mathbf{I}_{P}\right)^{-1}.
\end{equation}
The achievable EMSE $\EMSE(\mathbf{B})=\min_{\mathbf{D}}\mathbb{E}\{\|\tilde{\mathbf{s}}-\hat{\mathbf{s}}\|^2\}$ is
\begin{align}
	\EMSE(\mathbf{B})\!=\!\Tr\bigg[ \mathbf{M}\mathbf{P}^T\mathbf{R}_{c}&\left(\CovMat^{-1}\!-\!\mathbf{B}^H\left(\mathbf{B}\CovMat\mathbf{B}^H\!+\!\frac{4\gamma^2}{3b^2}\mathbf{I}_{P}\right)^{-1}\!\mathbf{B}\right) \notag \\
	&\times \mathbf{R}_{c}\mathbf{P}\mathbf{M}^H \bigg].
	\label{eqn:EMSE}
\end{align}
\end{lem}

\begin{IEEEproof}
The lemma is obtained as a special case of \cite[Lem. 1]{Shlezinger-TSP2019}, and thus we omit the proof for brevity.
\end{IEEEproof}

\smallskip
Using Lemma \ref{lemma1}, we optimize the analog filter matrix $\mathbf{B}$, which also dictates the support of the quantizers via \eqref{eqn_gamma}. We do so by designing $\mathbf{B}$ to minimize the EMSE $\EMSE(\mathbf{B})$ in \eqref{eqn:EMSE}, yielding the  filter  $\mathbf{B}^o$ given in the following theorem.

\begin{thm}
\label{theorem1}
    Let $\{\lambda_{\LMMSEMatW,l}\}$ be the singular values of $\LMMSEMatW\triangleq\LMMSEMat\CovMat^{1/2}$ arranged in a descending order.
	The analog combiner $\mathbf{B}^o$ which minimizes \eqref{eqn:EMSE} is given by $\mathbf{B}^o=\mathbf{U}^{o}\boldsymbol\Lambda^o(\mathbf{V}^o)^H\CovMat^{-1/2}$,
	where $\mathbf{V}^o$ is the right singular vectors matrix of $\LMMSEMatW$,
	$\boldsymbol\Lambda^o$ is a diagonal matrix with its diagonal entries given as
	\begin{equation}\label{eqn_optLambda}
	(\boldsymbol\Lambda^o)^2_{l,l}=\begin{cases}
		\frac{4\eta^2}{3b^2P}\left(\zeta\lambda_{\LMMSEMatW,l}-1\right)^{+}, &l\leq\min\{J,P\},\\
		0, &l>\min\{J,P\},
	\end{cases}
    \end{equation}
    and $\mathbf{U}^{o}$ is a unitary matrix such that $\mathbf{B}^o\CovMat(\mathbf{B}^o)^H=\mathbf{U}^{o}\boldsymbol\Lambda^o(\boldsymbol\Lambda^o)^T(\mathbf{U}^{o})^H$ has identical diagonal entries.
    In (\ref{eqn_optLambda}), $\zeta>0$ is set such that $\frac{4\eta^2}{3b^2P}\sum_{l=1}^{P}\left(\zeta\lambda_{\LMMSEMatW,l}-1\right)^{+}=1$.	
\end{thm}

\begin{IEEEproof}
\ifproofs
\else
The theorem follows from \cite[Thm. 1]{Shlezinger-TSP2019}.
\fi
\end{IEEEproof}

\smallskip
The unitary matrix $\mathbf{U}^{o}$  in Theorem \ref{theorem1} can be obtained via \cite[Alg. 2.2]{Palomar-Major}. 
With the  analog combining matrix $\mathbf{B}^o$, we can derive the EMSE and the resulting support of the quantizers.

\begin{corollary}
\label{cor:EMSEL1}
	For the \ac{bilimo} receiver with the analog combining matrix $\mathbf{B}^o$ given in Theorem \ref{theorem1}, the  quantizer support is $\gamma=\frac{\eta}{\sqrt{P}}$.
	The resulting achievable EMSE is given by 
	\begin{equation}\label{eqn_EMSEBo}
    \EMSE_o\!=\!\begin{cases}
	\sum\limits_{l=1}^{J}\frac{\lambda_{\LMMSEMatW,l}^2}{\left(\zeta\lambda_{\LMMSEMatW,l}-1\right)^{+}+1}, & P\!\geq \!J\\
	\sum\limits_{l=1}^{P}\frac{\lambda_{\LMMSEMatW,l}^2}{\left(\zeta\lambda_{\LMMSEMatW,l}-1\right)^{+}\!+\!1}\!+\!\sum\limits_{l=P\!+\!1}^{J} \lambda_{\LMMSEMatW,l}^2, &P\!<\! J.
\end{cases}
\end{equation}

\end{corollary}

\begin{IEEEproof}
	The dynamic range in (\ref{eqn_gamma}) is given by $\gamma^2=\frac{\eta^2}{P}\Tr(\boldsymbol\Lambda_o\boldsymbol\Lambda_o^T)=\frac{\eta^2}{P}$.
	The resulting EMSE in \eqref{eqn:EMSE} can be written as $\EMSE_o = \EMSE(\mathbf{B}^o)$ which is given by
\ifFullVersion
\begin{align}
		\EMSE_o&=\Tr\left[\LMMSEMatW\LMMSEMatW^H\right]-\sum_{l=1}^{\min\{J,P\}}\lambda_{\LMMSEMatW,l}^2 \frac{\left(\zeta\lambda_{\LMMSEMatW,l}-1\right)^{+}}{\left(\zeta\lambda_{\LMMSEMatW,l}-1\right)^{+}+1}.
\end{align}
\else
$\EMSE_o=\Tr\left[\LMMSEMatW\LMMSEMatW^H\right]-\sum_{l=1}^{\min\{J,P\}}\lambda_{\LMMSEMatW,l}^2 \frac{\left(\zeta\lambda_{\LMMSEMatW,l}-1\right)^{+}}{\left(\zeta\lambda_{\LMMSEMatW,l}-1\right)^{+}+1}$.
\fi
which coincides with \eqref{eqn_EMSEBo}. 
%
\end{IEEEproof}

\smallskip
The characterization of the \ac{bilimo} receiver configuration in Theorem \ref{theorem1} and the corresponding  accuracy in Corollary \ref{cor:EMSEL1} are obtained by expressing the problem of recovering the desired compressed representation as a task-based quantization setup \cite{Shlezinger-TSP2019}. This follows since for monotone waveforms, i.e., $L=1$, the effect of the analog filters $\{b_{p,n}(t)\}$ in \eqref{eqn_yp2} can be expressed as the matrix  $\mathbf{B}$, without imposing any structure constraints on the equivalent analog combining matrix. However, radar applications commonly utilize multitone signals, resulting in $L > 1$, which yields an equivalent formulation in which the analog combiner is constrained to take a structured form. Therefore, in the following we characterize the configuration of  \ac{bilimo} under such structure constraints.

\vspace{-0.2cm}
\subsection{\ac{bilimo} Receiver Design for Multitone Waveforms}
\label{subsec:Design2}
\vspace{-0.1cm}

For $L>1$, the analog combining matrix, which represents  the analog filters $\{b_{p,n}(t)\}$, is expressed as the product of $\bar{\mathbf{F}}$ and a block diagonal matrix $\bar{\mathbf{B}}$ as shown in Subsection~\ref{subsec:APP}.
By repeating the derivation in Lemma \ref{lemma1} with fixed analog processing, the  digital  filter  which  minimizes  the  MSE  for  a  fixed analog combiner $\bar{\mathbf{B}}$ is stated in the following lemma.

\begin{lem}\label{lemma2} 
For any analog combining matrix $\bar{\mathbf{B}}$, the digital processing matrix for the quantized output $\mathbf{z}$ in \eqref{eqn_model} which minimizes the MSE is given by
 \begin{equation}
	\mathbf{D}^\circ(\bar{\mathbf{B}}) = \mathbf{M}\mathbf{P}^T\mathbf{R}_{c}\bar{\mathbf{B}}^H\left(\bar{\mathbf{B}}\CovMat\bar{\mathbf{B}}^H+\frac{4\gamma^2}{3b^2}\mathbf{I}_{LP}\right)^{-1}\bar{\mathbf{F}}^H.
\end{equation}
The achievable EMSE $\EMSE(\bar{\mathbf{B}})=\min_{\mathbf{D}}\mathbb{E}\{\|\tilde{\mathbf{s}}-\hat{\mathbf{s}}\|^2\}$ is
\begin{align}
	\EMSE(\bar{\mathbf{B}})=\Tr\bigg[ \mathbf{M}\mathbf{P}^T\mathbf{R}_{c}&\left(\CovMat^{-1}\!-\!\bar{\mathbf{B}}^H  \left(\bar{\mathbf{B}}\CovMat\bar{\mathbf{B}}^H\!+\!\frac{4\gamma^2}{3b^2}\mathbf{I}_{LP}\right)^{-1}\!\bar{\mathbf{B}}\right) \notag \\
	&\times\mathbf{R}_{c}\mathbf{P}\mathbf{M}^H \bigg].
	\label{eqn_EMSE_barB}
\end{align}
\end{lem}
\begin{IEEEproof}
Substituting $\bar{\mathbf{F}}\bar{\mathbf{B}}$ for $\mathbf{B}$ in the derivation of Lemma~\ref{lemma1}, and applying the relation $\bar{\mathbf{F}}\bar{\mathbf{F}}^H=\bar{\mathbf{F}}^H\bar{\mathbf{F}}=\mathbf{I}_{LP}$, proves the lemma.
\end{IEEEproof}

\smallskip
The derivation of the digital processing for a given analog filter is invariant to whether the waveforms are monotone or multitone. However, when optimizing the analog combining matrix in light of \eqref{eqn_EMSE_barB}, one must account for its block-diagonal structure induced when $L >1$. In particular, we design the matrix  $\bar{\mathbf{B}}$ which accounts for the aforementioned constraint by formulating the following objective:
\begin{equation}\label{prob_EMSEdiag} 
    	\min_{\mathbf{B}_1,\cdots,\mathbf{B}_L\in\mathbb{C}^{P\times MN}} \EMSE(\bar{\mathbf{B}}=\blkdiag\{\mathbf{B}_1,\cdots,\mathbf{B}_L\}).
\end{equation}
In order to tackle \eqref{prob_EMSEdiag}, we can choose an appropriate matrix $\mathbf{M}$ such that the matrix $\mathbf{M}\mathbf{P}^H$ is also block diagonal. Let $\mathbf{M}_i$ be the $i$th block of $\mathbf{M}\mathbf{P}^H$ with its dimension $J_i\times MN$, where $\sum_{i=1}^LJ_i=J$. Then the compressed vector $\mathbf{s}$ can be separated into $L$ sub-vectors, each related to a different vector in the set $\{\mathbf{c}_i\}$, i.e., $\mathbf{s}_i=\mathbf{M}_i\mathbf{c}_i$. Furthermore, we introduce the following assumption on the underlying statistical model: 
\begin{itemize}
     \item [A1] The covariance matrices of $\mathbf{c}$ and $\mathbf{w}$, i.e., $\mathbf{R}_{c}$ and $\mathbf{R}_w$, are block diagonal, with their $i$-th blocks being the $MN\times MN$ matrices  $\mathbf{R}_{c_i}=\mathbb{E}\{\mathbf{c}_i\mathbf{c}_i^H\}$ and $\mathbf{R}_{w_i}$, respectively.
\end{itemize}
Assumption A1 means that we only consider the correlation within the vector $\mathbf{c}_i$ and ignore the correlation between the different vectors $\mathbf{c}_i$ and $\mathbf{c}_{i'}$ for $i\neq i'$. Since each $\mathbf{c}_i$ represents the samples of the received echos after channel separation at a given tone index $i$ by \eqref{eqn_cmni}, A1 implies that echos observed at different frequency bins are uncorrelated.The validity of this assumption clearly depends on the distribution of $\{\alpha_k\}$, $\{\tau_k\}$, and $\{\varphi_k\}$. As shown in \cite{Eldar2015SPL}, if $\alpha_k$, $\tau_k$, and $\varphi_k$ are mutually independent, with $\{\alpha_k\}$ being zero-mean i.i.d. and $\tau_k$ and $\varphi_k$ being uniformly distributed on $[0, T_0]$ and $[-1,1]$, respectively, then $\{c_{m,n}[i]\}$ are uncorrelated. In this case, the covariance matrix $\mathbf{R}_c$ is diagonal, satisfying  assumption A1.
With this assumption, $\CovMat$ also becomes a block diagonal matrix, with its $i$-th block being the $MN\times MN$ matrix $\CovMat_i=\mathbf{R}_{c_i}+\mathbf{R}_{w_i}$.

Under  assumption A1,  we characterize the block diagonal matrix $\bar{\mathbf{B}}^o$ which solves \eqref{prob_EMSEdiag}  in the following theorem:
\begin{thm}
\label{theorem2}
    Let $\{\lambda_{\LMMSEMatW,l}^{(i)}\}$ be the singular values of $\LMMSEMatW_i\triangleq\mathbf{M}_i\mathbf{R}_{c_i}\CovMat^{-1/2}_i$ arranged in a descending order.
	Each block of the  block diagonal matrix $\bar{\mathbf{B}}^o=\blkdiag\{\mathbf{B}_1^o,\cdots,\mathbf{B}_L^o\}$  which solves \eqref{prob_EMSEdiag} is given by $\mathbf{B}_i^o=\mathbf{U}^{o}_i\boldsymbol\Lambda^o_i(\mathbf{V}^o_i)^H\CovMat^{-1/2}_i$,
	where $\mathbf{V}_i^o$ is the right singular vectors matrix of $\LMMSEMatW_i$,
	$\boldsymbol\Lambda_i^o$ is a diagonal matrix with its diagonal entries given as
	\vspace{-0.1cm}
	\begin{equation}\label{eqn_optLambdai}
	(\boldsymbol\Lambda_i^o)^2_{l,l}=\begin{cases}
		\frac{4\eta^2}{3b^2P}\left(\zeta_i\lambda_{\LMMSEMatW,l}^{(i)}-1\right)^{+}, &l\leq\min\{J_i,P\}\\
		0, &l>\min\{J_i,P\},
	\end{cases}
	\vspace{-0.1cm}
    \end{equation}
    and $\mathbf{U}^{o}_i$ is a unitary matrix such that $\mathbf{B}^o_i\CovMat_i(\mathbf{B}^o_i)^H=\mathbf{U}^o_i\boldsymbol\Lambda^o_i({\boldsymbol\Lambda}^o_i)^T(\mathbf{U}^o_i)^H$ has identical diagonal entries. 
    In \eqref{eqn_optLambdai}, $\zeta_i>0$ is set such that $\frac{4\eta^2}{3b^2P}\sum_{l=1}^{P}\left(\zeta_i\lambda_{\LMMSEMatW,l}^{(i)}-1\right)^{+}=1$.	
\end{thm}
\begin{IEEEproof}
The proof is given in Appendix \ref{app:Proof2}.
\end{IEEEproof}

\smallskip
The analog combiner $\bar{\mathbf{B}}^o$ in Theorem~\ref{theorem2}, 
is obtained by optimizing the individual contribution of each spectral component indexed by $i \in \{1,\ldots, L\}$, using the results obtained for the monotone case in Theorem \ref{theorem1} for each spectral component. In general, the optimization problem in \eqref{prob_EMSEdiag} cannot be immediately converted into $L$ individual problems, since the ADCs have a fixed support $\gamma$ which depends on overall analog combiner matrix $\bar{\mathbf{B}}^o$, and thus the problems are inherently coupled. Nonetheless, as we show in Appendix  \ref{app:Proof2}, one can still apply the monotone design of Theorem \ref{theorem1} for each spectral component separately, as the combination of the matrix $\bar{\mathbf{F}}$ and the unitary matrices $\{\mathbf{U}_i^o\}$ in  Theorem \ref{theorem2} result in each spectral component having the same effect on the setting of  $\gamma$.

With the optimal block diagonal matrix $\bar{\mathbf{B}}_o$ given in Theorem \ref{theorem2}, we can derive the optimal dynamic range of the quantizers, as well as the resulting MSE.

\begin{corollary}
\label{cor:EMSEL2}
	For the \ac{bilimo} receiver with the block diagonal  $\bar{\mathbf{B}}^o$ given in Theorem \ref{theorem2}, the dynamic range of the quantizer is $\gamma=\frac{\eta}{\sqrt{P}}$.
	The resulting EMSE is $\EMSE_o=\sum_{i=1}^L \varepsilon_i$, where
	\vspace{-0.1cm}
	\begin{equation}\label{eqn_epsilon}
    \varepsilon_i=\begin{cases}
	\sum\limits_{l=1}^{J_i}\frac{(\lambda_{\LMMSEMatW,l}^{(i)})^2}{\left(\zeta\lambda_{\LMMSEMatW,l}^{(i)}-1\right)^{+}+1}, & P\geq J_i\\
	\sum\limits_{l=1}^{P}\frac{(\lambda_{\LMMSEMatW,l}^{(i)})^2}{\left(\zeta\lambda_{\LMMSEMatW,l}^{(i)}-1\right)^{+}+1}+\sum\limits_{l=P+1}^{J_i} (\lambda_{\LMMSEMatW,l}^{(i)})^2, &P<J_i.
\end{cases}
	\vspace{-0.1cm}
\end{equation}
\end{corollary}

\begin{IEEEproof}
	The dynamic range is given by $\gamma^2=\frac{\eta^2}{PL}\sum_{i=1}^L\Tr(\boldsymbol\Lambda^o_i(\boldsymbol\Lambda^o_i)^T)=\frac{\eta^2}{P}$.
	The   EMSE is $\EMSE_o=\sum_{i=1}^L \varepsilon_i$, where 
\ifFullVersion
\begin{align} 
	    \varepsilon_i&=\Tr\left(\mathbf{T}_i\CovMat_i^{-1}\mathbf{T}_i^H\right)-\sum_{l=1}^{\min\{J_i,P\}}(\lambda_{\LMMSEMatW,l}^{(i)})^2 \frac{\left(\zeta\lambda_{\LMMSEMatW,l}^{(i)}-1\right)^{+}}{\left(\zeta\lambda_{\LMMSEMatW,l}^{(i)}-1\right)^{+}+1} \notag \\
		&=\sum_{l=1}^{J_i}(\lambda_{\LMMSEMatW,l}^{(i)})^2-\sum_{l=1}^{\min\{J_i,P\}}(\lambda_{\LMMSEMatW,l}^{(i)})^2 \frac{\left(\zeta\lambda_{\LMMSEMatW,l}^{(i)}-1\right)^{+}}{\left(\zeta\lambda_{\LMMSEMatW,l}^{(i)}-1\right)^{+}+1}.
\label{eqn_epsilon1}
\end{align}
By simplifying the above expression, we obtain \eqref{eqn_epsilon}.
\else
    $\varepsilon_i=\Tr\big(\mathbf{T}_i\CovMat_i^{-1}\mathbf{T}_i^H\big)-\sum_{l=1}^{\min\{J_i,P\}}(\lambda_{\LMMSEMatW,l}^{(i)})^2 \frac{\left(\zeta\lambda_{\LMMSEMatW,l}^{(i)}-1\right)^{+}}{\left(\zeta\lambda_{\LMMSEMatW,l}^{(i)}-1\right)^{+}+1}$. 
    As $\Tr\big(\mathbf{T}_i\CovMat_i^{-1}\mathbf{T}_i^H\big) = \sum_{l=1}^{J_i}(\lambda_{\LMMSEMatW,l}^{(i)})^2$, simplifying this expression proves \eqref{eqn_epsilon}.
\fi
%
\end{IEEEproof}


\vspace{-0.2cm}
\subsection{Target Reconstruction by Sparse Recovery}
\label{subsec:Analysis}
\vspace{-0.1cm}
Using the characterized \ac{bilimo} receivers detailed in the previous subsections, we can acquire an estimate of $\mathbf{s}$, i.e., $\hat{\mathbf{s}}$, which minimizes the MSE between the estimate $\hat{\mathbf{s}}$ and the LMMSE estimate $\tilde{\mathbf{s}}$.
This allows the radar receiver to obtain an accurate estimate of $\mathbf{s}$, which is a compressed representation of the targets range-delay grid vector $\TarVec$, as we also numerically demonstrate in our simulation study in Section \ref{sec:Sims}.
Having obtained the compressed representation of the targets grid $\hat{\mathbf{s}}$, the task of recovering the targets information can be formulated as the recovery of the $K$-sparse vector $\TarVec$ from $\hat{\mathbf{s}}$. Following  sparse recovery methods \cite[Ch. 1]{eldar2012compressed}, this task can be relaxed into the following $\ell_1$ minimization:
	\vspace{-0.1cm}
\begin{equation}
    \label{Prob:SparseRec}
     \hat{\TarVec} = \min_{\mathbf{a}}  \|\mathbf{a}\|_1
     \quad \text{s.t.} \quad \|\hat{\mathbf{s}}-\mathbf{M}\mathbf{\Phi}\mathbf{a}\|_2^2\leq \tilde{\varepsilon}, 
	\vspace{-0.1cm}
\end{equation}
or the LASSO problem
	\vspace{-0.1cm}
\begin{equation}
    \label{Prob:Lasso}
    \hat{\TarVec} = \min_{\mathbf{a}} \frac{1}{2}\|\hat{\mathbf{s}}-\mathbf{M}\mathbf{\Phi}\mathbf{a}\|_2^2 + \rho \|\mathbf{a}\|_1, 
	\vspace{-0.1cm}
\end{equation}
where $\tilde{\varepsilon}$ and $\rho$ are predeﬁned regularization parameters.
The optimization problems \eqref{Prob:SparseRec} and  \eqref{Prob:Lasso} can be conveniently solved by convex optimization methods, such as FISTA  \cite{Beck2009FISTA}, which we use in our numerical study to solve \eqref{Prob:Lasso}.
We can also utilize matrix-form sparse recovery algorithms, as done in \cite{Cohen2018-TSP}, by exploiting the structure of $\mathbf{\Phi}$. 
This procedure allows the \ac{bilimo} receiver to mitigate the effect of the limited bit budget on its ability to recover the targets. 
This is achieved by tuning the system to recover a compressed representation, rather than processing the high-dimensional  echos in digital, thus mitigating the quantization distortion while maintaining the ability to reconstruct the targets in the digital domain.

The fact that the \ac{bilimo} receiver applies sparse recovery methods implies that its reconstruction error can be analytically bounded using results from \ac{cs} theory.
Therefore, we next show this bounding the error when solving the sparse recovery problem \eqref{Prob:SparseRec}. To that aim, we recall the definition of the matrix coherence measure, commonly used in sparse recovery analysis.  The coherence of a matrix $\mathbf{A}$, $\mu(\mathbf{A})$, is the largest absolute inner product between any two columns $\mathbf{a}_i$, $\mathbf{a}_j$ of $\mathbf{A}$, i.e., $\mu(\mathbf{A})=\max_{i,j}\frac{|<\mathbf{a}_i,\mathbf{a}_j>|}{\|\mathbf{a}_i\|_2\|\mathbf{a}_j\|_2}$. By letting $\LMMSE$ be the LMMSE, i.e., $\LMMSE \triangleq \mathbb{E}\{\|\mathbf{s} - \tilde{\mathbf{s}}\|^2_2\}$, we can bound the targets reconstruction error as stated in the following theorem:
\begin{thm}
\label{theorem3}
When the quantizers are not overloaded and the number of targets satisfies $K<(1/\mu(\mathbf{M}\mathbf{\Phi})+1)/4$, then the proposed \ac{bilimo} receiver  recovers the targets vector $\hat{\TarVec}$ via \eqref{Prob:SparseRec} 
within an error which is bounded by
	\vspace{-0.1cm}
\begin{equation}
  \mathbb{E}\{\|\TarVec-\hat{\TarVec}\|_2^2 \}\leq \frac{\LMMSE + \EMSE_o + \tilde{\varepsilon}}{1-(4K-1)\mu(\mathbf{M}\mathbf{\Phi})}.
  \label{eqn:theorem3}
	\vspace{-0.1cm}
\end{equation}
\end{thm}

\begin{IEEEproof}
As shown earlier, when the quantizers are not overloaded,  \ac{bilimo}  achieves an EMSE of $\EMSE_o =   \mathbb{E}\{\|\tilde{\mathbf{s}}- \hat{\mathbf{s}}\|^2_2\}$. In such a case, we can write  $ \hat{\mathbf{s}} = \mathbf{s} + \mathbf{e}$ where $\mathbf{e}$ is the overall error satifying
\ifFullVersion
\begin{equation}
\label{eqn:Error}
    \mathbb{E}\{\|\mathbf{e}\|^2_2\}= \mathbb{E}\{\|({\mathbf{s}} \! - \tilde{\mathbf{s}}) \! + \! (\tilde{\mathbf{s}}\!-\!\hat{\mathbf{s}}) \|^2_2\} \stackrel{(a)}{=} \LMMSE + \EMSE_o.
\end{equation}
\else
$   \mathbb{E}\{\|\mathbf{e}\|^2_2\}= \mathbb{E}\{\|({\mathbf{s}} \! - \tilde{\mathbf{s}}) \! + \! (\tilde{\mathbf{s}}\!-\!\hat{\mathbf{s}}) \|^2_2\} \stackrel{(a)}{=} \LMMSE + \EMSE_o$.
\fi
Here, $(a)$ follows from the orthogonality principle combined with the fact that $\hat{\mathbf{s}}$ obtained by task-based quantization with dithered quantizers can be modeled as a linear function of $\mathbf{c}$  corrupted by additive uncorrelated noise \cite[Lem. 1]{Shlezinger-TSP2019}. Combining this with the stability bound for $\ell_1$-minimization based sparse recovery in \cite[Theorem 1.11]{eldar2012compressed} proves \eqref{eqn:theorem3}.
\end{IEEEproof}

\smallskip
%
%
Theorem \ref{theorem3} bounds the achievable error in recovering the target parameters by the proposed \ac{bilimo} receiver. The result also holds for sparse recovery with arbitrary value of $\varepsilon$, without imposing any limits on choosing the predefined parameter. In particular, for a given number of targets $K$, the bound is proportional to the overall MSE, which is comprised of two terms: the LMMSE $\LMMSE$, that is an inherent property of the signal model and is a byproduct of fact that the echos are noisy; and the EMSE $\EMSE_o$, which follows from the bit constraints. This result  implies that the target recovery  performance of the proposed \ac{bilimo} receiver, which is designed to minimize the EMSE due to quantization constraints, is not expected to achieve perfect recovery due to the inherent error induced by the presence of noise. Nonetheless, we are interested in the sparsity pattern of $\mathbf{a}$, from which the delays and angles can be extracted, rather than its actual values. Consequently, by mitigating the distortion due to quantization, the \ac{bilimo} receiver is capable of approaching ideal recovery at signal-to-noise ratio (SNR) values as low as $-10$ dB, while operating under tight bit budgets equivalent to one bit per sample, as demonstrated in our numerical study presented in Section \ref{sec:Sims}. 

The error bound in Theorem \ref{theorem3} is 
inversely proportional to the coherence of  $\mathbf{M}\mathbf{\Phi}$. This  provides us some guidelines to determine the measurement matrix $\mathbf{M}$. The setting of $\mathbf{M}$ should account for two key considerations: First, its number of rows $J$ should satisfy $J \leq P L$, as noted in Subsection~\ref{subsec:problem}. The value of $P$ should not be larger than  $MN$, and reducing $P$  implies that we are using less ADCs, and can thus allocate more bits to each quantizer under a given bit budget, and thus one would wish to use small values of $J$. However, the resulting compressed representation should also be sufficient to allow recovering the desired target parameters grid $\mathbf{a}$. By Theorem \ref{theorem3}, this is achieved by setting $\mathbf{M}$ such that the coherence measure $\mu(\mathbf{M}\mathbf{\Phi})$ satisfies $K<(1/\mu(\mathbf{M}\mathbf{\Phi})+1)/4$, preferably using as small  coherence  as possible. 
In our numerical study  in Section \ref{sec:Sims}, we generate the entries of  $\mathbf{M}$ from a  complex Gaussian distribution, and set it to recover a  representation which is smaller by factors of $2$ and $4$ compared to the number of elements in $\mathbf{\Phi}\mathbf{a}$ (which equals $\mathbf{s}$ when $\mathbf{M} = \mathbf{I}_{J}$). This setting is numerically shown to yield reliable target identification under tight bit constraints, allowing to approach perfect recovery of the target parameters for SNR  larger than $-10$ dB while utilizing no more than one bit per input sample. 



\vspace{-0.2cm}
\subsection{Discussion}
\label{subsec:Discussion}
\vspace{-0.1cm}
The proposed \ac{bilimo} receiver exploits the task for which the  echos are acquired  to facilitate target detection under bit constraints. The signals acquired by MIMO radar receivers are typically high-dimensional, and thus require a large bit budget to be converted into a digital representation in a manner which allows their reconstruction.    Our task-based design builds upon the  insight that the receiver is interested in the target parameters rather than reconstructing the  echos, and that the targets can still be accurately recovered from a coarse low-resolution quantized version of the measurements. 
Therefore, the radar receiver task can be treated as indirect lossy source coding of compressed measurements \cite{leinonen2018rate}, in which the system is the HAD receiver detailed in Section \ref{sec:System}.
The \ac{bilimo} receiver  accounts for this task by  designing the analog filters $\{b_{p,n}(t)\}$ such that the inputs to the uniform \acp{adc} maintain approximate sufficiency with respect to the desired information, i.e., the compressed vector $\mathbf{s}$, while resulting in a minor level of distortion induced in quantization. In particular, the waterfilling-type expression in \eqref{eqn_optLambda} and \eqref{eqn_optLambdai} preserves the dominant eigenmodes of the LMMSE estimate of $\mathbf{s}$, and nullifies the weak ones which become indistinguishable in uniform quantization. The unitary matrix $\mathbf{U}^o$ guarantees that each \ac{adc} quantizes an input with the same variance, allowing to minimize the maximal quantization distortion. This operation effectively balances the ability to estimate $\mathbf{s}$ from the \acp{adc} analog input along with the distortion induced in quantization in light of the overall system task. 

The common strategy to design radar receivers operating under bit constraints is to carry out recovery in the digital domain based on low-resolution  measurements, i.e., without analog combining, as in \cite{xi2020joint}. Alternatively, in the presence of controllable analog processing, an intuitive approach is to design the analog filter to  estimate  ${\mathbf{s}}$ as accurately as possible. This approach is known to minimize the overall MSE when using vector quantizers \cite{wolf1970transmission}.  In the presence of scalar uniform \acp{adc}, as commonly utilized in radar applications, HAD systems such as  \ac{bilimo}  designed in a task-based manner were proven to outperform the aforementioned approaches when the  task is  a linear function of the measurements \cite{Shlezinger-TSP2019}. 
While we design the receiver to recover the  compressed    $\mathbf{s}=\mathbf{M}\mathbf{\Phi}\TarVec$ only as an intermediate step in identifying the targets, Theorem \ref{theorem3} proves that accurately estimating $\mathbf{s}$ directly affects  target recovery. Furthermore, in Section \ref{sec:Sims} we numerically demonstrate that designing a hybrid system to estimate $\mathbf{s}$ under bit constraints yields improved accuracy in recovering the targets over purely digital strategies, as well as approaches the accuracy achievable without quantization constraints. 

 \ac{bilimo}  illustrated in Fig. \ref{weightfunction} implements task-based quantization  by introducing pre-acquisition analog filtering. 
Since the design of this analog filter depends on the underlying statistical model of the  echos, realizing such a  receiver  requires using controllable analog combiner hardware. Such combining which can realize various forms of analog filters can be implemented using dedicated circuitry as proposed in \cite{gong2020rf}. An alternative approach, which may be preferable in some applications, is to implement analog combining via phase-shifter networks \cite{mendez2016hybrid,ioushua2019family}, or using externally configurable antennas \cite{Shlezinger-TOC2019,shlezinger2020dynamic}. Such architectures induce some constraints on the  feasible analog filter, which follow from their specific hardware, as in, e.g. \cite{Wang-2019DMA}. We leave the analysis and design of \ac{bilimo} with constrained analog filters for future work.

Our design of  \ac{bilimo}  requires prior knowledge of the underlying statistical model, and particularly the covariance matrices of $\mathbf{c}$ and  $\mathbf{w}$, denoted $\mathbf{R}_c$ and $\mathbf{R}_w$, respectively. While the noise matrix $\mathbf{R}_w$ can be typically assumed to be some scalar multiple of the identity matrix, representing i.i.d. measurement noise, obtaining $\mathbf{R}_c$ may be challenging.
One possible way is to exploit some a prior distribution of the targets, which is often known to the receiver based on some pre-training, to  estimate the covariance matrix $\mathbf{R}_c$. 
An additional quantity which is required in sparse recovery is the number of targets $K$.
This problem is equivalent to the model order selection problem, which can be solved by utilizing the Bayesian information criterion.
Finally, one can also overcome the need to know the underlying statistical model by designing the components of  \ac{bilimo}  in a data-driven manner, using machine learning methods for tuning task-based quantizers  \cite{shlezinger2019deep,shlezinger2020learning}. We leave the design of \ac{bilimo} receivers with partial and missing model knowledge to future work.

\vspace{-0.2cm}
\section{Numerical Results}
\label{sec:Sims}
\vspace{-0.1cm}
In this section, we present numerical experiments illustrating the performance of  \ac{bilimo}. 
We compare our method with MIMO radar systems without quantization constraints as well as with digital MIMO radar receivers which digitize the signal after channel separation via matched filtering  with the same bit budget, i.e., using task-ignorant quantizers, and then identifies the targets via sparse recovery using the FISTA algorithm.

\vspace{-0.2cm}
\subsection{Simulation Setup}
\label{subsec:Simsetup}
\vspace{-0.1cm}
Throughout the simulations, we consider a MIMO radar with $M=8$ transmit antennas and $N=12$ receive antennas.
The locations of the antennas are uniformly randomized over the virtual aperture $MN\lambda/2$, as done in \cite{Cohen2018-TSP}.
We use a set of multitone waveforms $h_m(t)$ such that $f_m = (i_m-\frac{M+1}{2})B_h$, 
where $i_m$ are integers chosen uniformly at random in $[0,M)$.
The remaining simulation parameters are: PRI $T_0=9$ $\mu$sec, bandwidth $B_h = 1 \text{MHz}$, and carrier frequency $f_c= 10 \text{GHz}$.
Accordingly, the value of $L$ can be derived as $L=9$.
The parameters $\tau_k$ and $\vartheta_k$ of each target are randomly generated on the delay-angle grids defined in Subsection \ref{subsec:DP} with grid spacing $\Delta_{\tau}=\frac{T_0}{72}$ and $\Delta_{\vartheta}=\frac{2}{96}$, respectively, i.e., $\tau_k\in \{0,\Delta_{\tau},2\Delta_{\tau},\cdots,T_0\}$ and $\vartheta_k\in \{-1,-1+\Delta_{\vartheta},-1+2\Delta_{\vartheta},\cdots,1\}$.
The received signals are corrupted with i.i.d. additive proper-complex Gaussian noise with zero mean and variance $\sigma_n^2$. 
We define the SNR as 
   $ \text{SNR} = \frac{\mathbb{E}\{\|\mathbf{\Phi}\mathbf{a}\|_2^2\}}{MNLK\sigma_n^2}$.

\begin{figure}
\centering
\includegraphics[width=\figWidth, height=\figHeight]{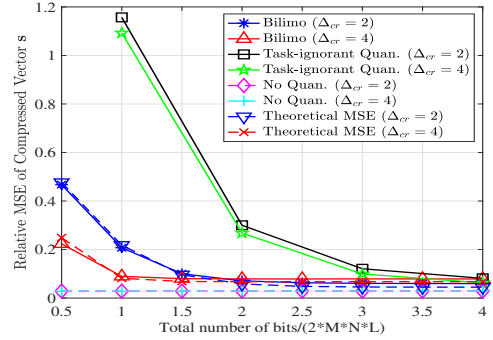}
\vspace{-0.1cm} 
\caption{MSE in recovering $\mathbf{s}$ versus total number of bits (SNR=10dB).}
\label{MSEvsBits_s}
\end{figure}

In the \ac{bilimo} receiver, we define the compression ratio as $\Delta_{cr}=\frac{MNL}{J}$ to evaluate the impact of the dimension of the compressed vector $\mathbf{s}$.
Each block of the matrix $\mathbf{M}\mathbf{P}^H$ is a complex Gaussian random matrix, with its entries being i.i.d. circularly-symmetric complex Gaussian random variables with zero mean and unit variance, if not specified.
The number of analog channels is set to be $P=\lceil J/L\rceil$.
In the MIMO radar with task-ignorant quantizers, the receiver first separates each channel from the received signals and then quantizes them with the same overall bit-budget, regardless of the task.
The quantized outputs are used to identify the targets via sparse recovery.
For the MIMO radar without quantization, two different methods are considered:
One is to directly recover the target vector $\TarVec$ from the received signal $\mathbf{y}+\mathbf{n}$; The second first computes the LMMSE estimate $\tilde{\mathbf{s}}$, and then recovers the target vector $\TarVec$ from the LMMSE estimate.
These two methods are denoted as ``No Quan. (DR)'' and ``No Quan. (LMMSE)'', respectively. 
For sparse recovery we use  FISTA  \cite{Beck2009FISTA} to solve the LASSO problem  \eqref{Prob:Lasso}.
We evaluate the various methods by repeating each experiment over 100 realizations.

\begin{figure}
\centering
\includegraphics[width=\figWidth, height=\figHeight]{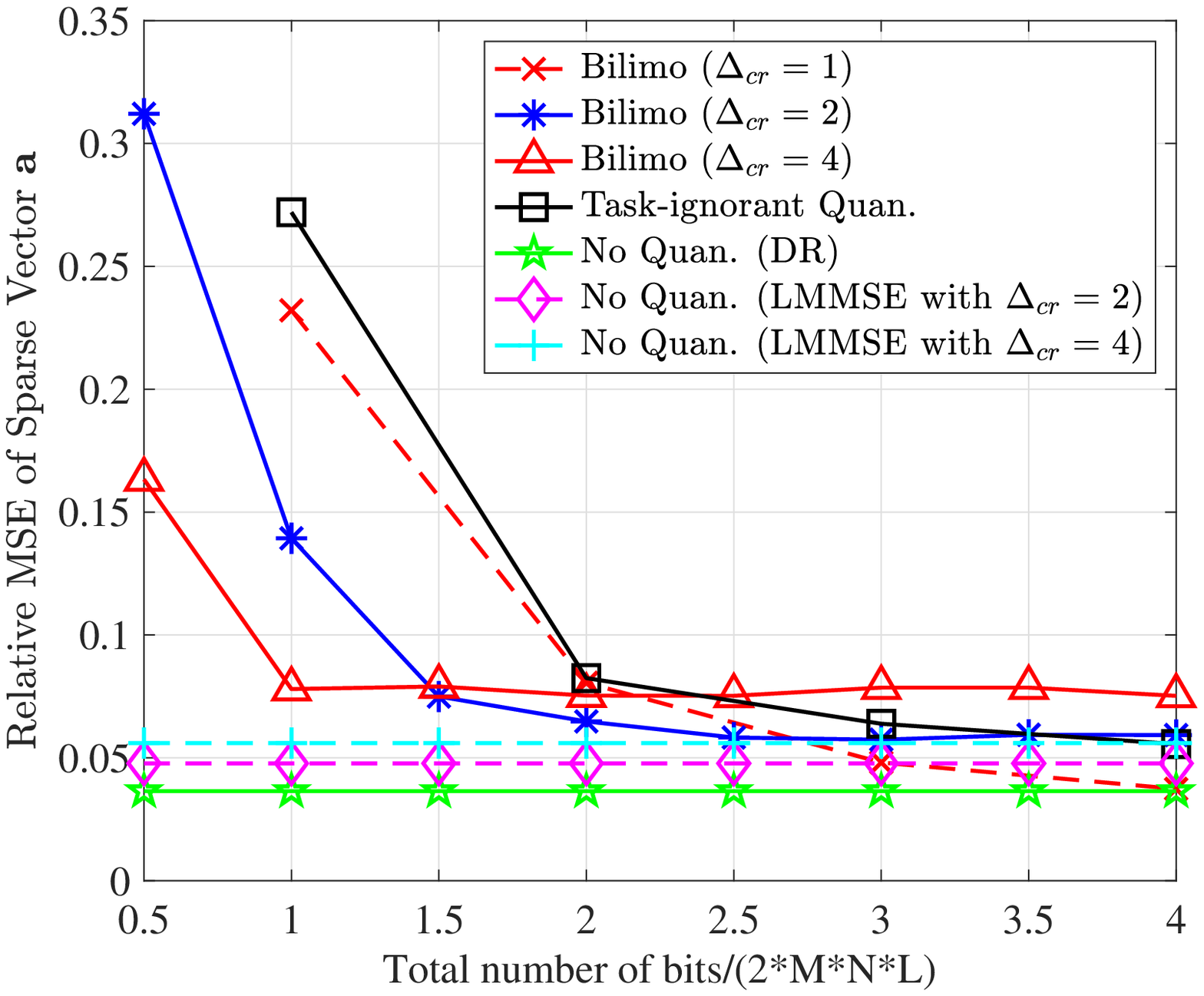} 
\vspace{-0.1cm}
\caption{MSE in recovering $\TarVec$ versus total number of bits (SNR=10dB).}
\label{MSEvsBits_a}
\end{figure}

\vspace{-0.2cm}
\subsection{Estimation Performance}
\label{subsec:EstiPerform}
\vspace{-0.1cm}
We begin by evaluating the estimation performance in recovering the target vector $\TarVec$, as well as the compressed vector $\mathbf{s}$.
The relative MSEs of $\mathbf{a}$ and $\mathbf{s}$, respectively defined as $\|\mathbf{a}-\hat{\mathbf{a}}\|_2^2/\|\mathbf{a}\|_2^2$ and $\|\mathbf{s}-\hat{\mathbf{s}}\|_2^2/\|\mathbf{s}\|_2^2$, are used as metrics.
We consider $K=4$ targets, and the targets reflection coefficients $\{\alpha_k\}$ are randomized as i.i.d. proper-complex Gaussian random variables with zero mean and unit variance. 

\begin{figure}
\centering
\includegraphics[width=\figWidth, height=\figHeight]{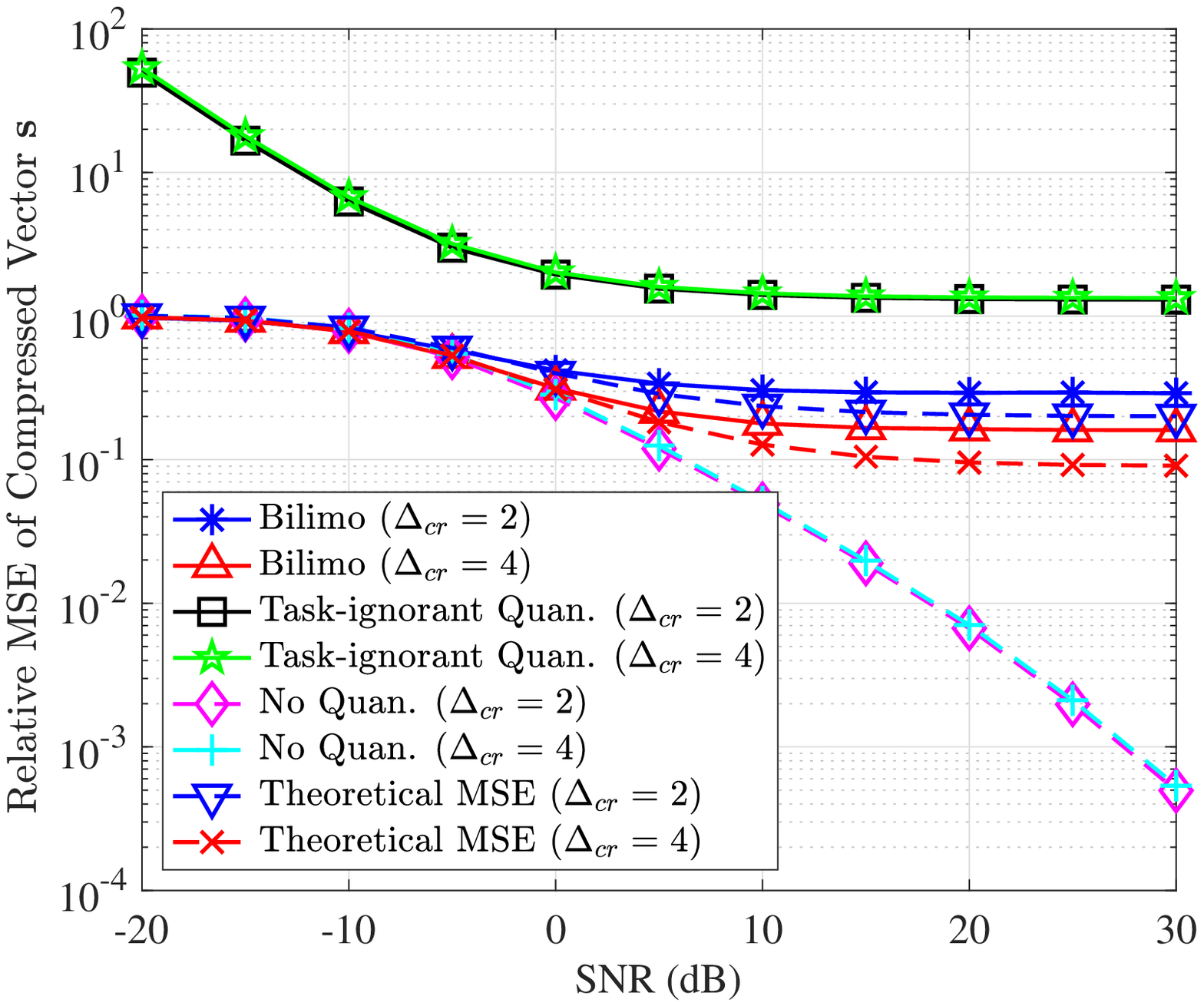} 
\vspace{-0.1cm}
\caption{MSE in recovering $\mathbf{s}$ versus SNR (Budget of $2MNL=1728$ bits).}
\label{MSEvsSNR_s}
\end{figure}

We first investigate the estimation performance versus the overall bit budget.
Fig.~\ref{MSEvsBits_s} depicts of the relative MSEs in recovering the compressed $\mathbf{s}$ with different compression ratio $\Delta_{cr}$.
To assert our theoretical MSE derivation in Theorem \ref{theorem2}, we also depict the theoretical performance of  \ac{bilimo}, given by the sum of the LMMSE and the resulting EMSE.
It is observed that  \ac{bilimo}  significantly outperforms the task-ignorant quantization operating with the same number of bits. Furthermore,  \ac{bilimo}  achieves MSE performance which is within  a small gap from the ``No Quan.'' methods, which operate with infinite resolution ADCs, while operating with as few as two bits per input sample. 
Comparing  \ac{bilimo}  with different values of $\Delta_{cr}$, we find that  \ac{bilimo}  with larger $\Delta_{cr}$ achieves more accurate representations of $\mathbf{s}$ in the low bit-budget regime, since more bits can be assigned to each ADC. However, for constraints of more than two bits per sample, using  the smaller compression of  $\Delta_{cr} = 2$ achieves improved performance. 
The theoretical curves closely effectively coincide the simulated curves, validating our theoretical analysis.

\begin{figure}
\centering
\includegraphics[width=\figWidth, height=\figHeight]{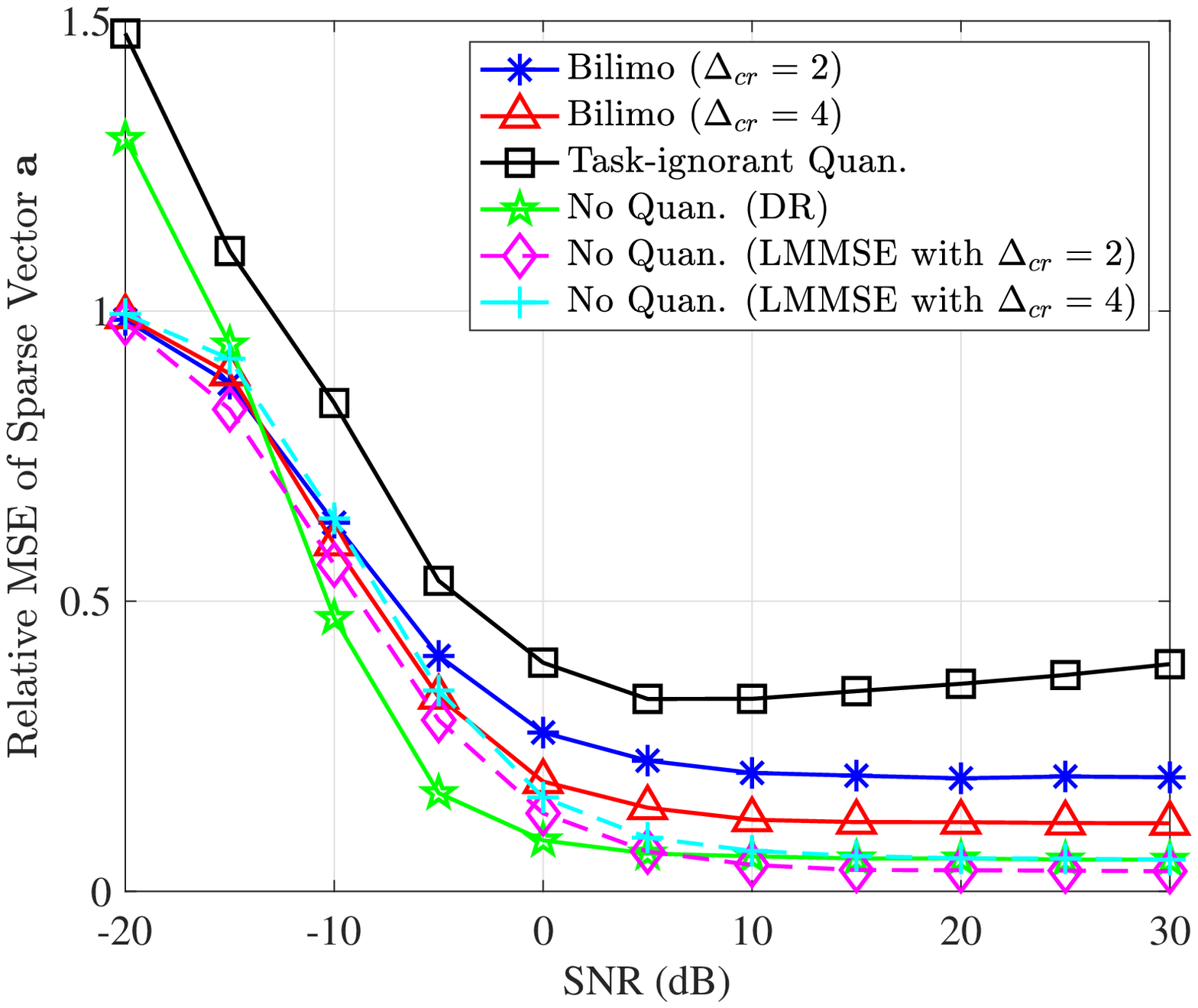} 
\vspace{-0.1cm}
\caption{MSE in recovering $\TarVec$ versus SNR (Budget of $2MNL=1728$ bits).}
\label{MSEvsSNR_a}
\end{figure}

In Fig. \ref{MSEvsBits_a}, the relative MSE in recovering the target vector $\TarVec$ with respect to different bit-budgets are shown.
From this figure, we see that the \ac{bilimo} receiver substantially outperforms the task-ignorant quantization when the bit-budget is low, e.g., when the total number of bits is less than twice the data dimension. 
It is also observed that using high compression ratio results in improved recovery at low quantization rates, as this compression allows to trade sufficiency for reduced quantization distortion by assigning more bits to each ADC without violating the overall bit constraint. As the overall number of bits increases, the errors induced due to compression become more notable compared to the quantization distortion, and lower compression ratios, e.g., $\Delta_{cr} \leq 2$, are preferred.  
This result is consistent with our theoretical result in Theorem \ref{theorem3} since the coherence $\mu(\mathbf{M}\mathbf{\Phi})$ becomes larger as the compression ratio increases.
In Fig. \ref{MSEvsBits_a}, we also depict the MSE curve of the \ac{bilimo} receiver with $\Delta_{cr}=1$, which achieves lower MSE values compare to task-ignorant quantization even in the high bit budget regime, demonstrating that properly designed analog combiners contribute to the overall performance even when they do not reduce the dimensionality of the inputs. 

Next, we investigate the estimation performance  versus the SNR for different compression ratios. The MSEs in recovering $\mathbf{s}$ and $\mathbf{a}$ are depicted in Figs. \ref{MSEvsSNR_s}-\ref{MSEvsSNR_a}, respectively.
Observing Figs. \ref{MSEvsSNR_s}-\ref{MSEvsSNR_a}, we note that the MSEs in recovering both $\mathbf{s}$ and $\TarVec$ are largely decreased by applying the \ac{bilimo} receiver, compared with  task-ignorant quantization.
For  SNRs lower than $-10$ dB,  \ac{bilimo}  approaches the performance of the unquantized ``No Quan. (LMMSE)'' method and achieves even better performance than the ``No Quan.(DR)'' method. This demonstrates that designing the quantization strategy based on the LMMSE estimator significantly decreases the effect of noise on the estimation.
As the SNR increases, the quantization distortion   becomes the dominant source of errors, and it is observed that while \ac{bilimo} notably outperforms conventional task-ignorant strategies, its error does not grow arbitrarily small as in the infinite quantization resolution case. 

\begin{figure}
\centering
\includegraphics[width=\figWidth, height=\figHeight]{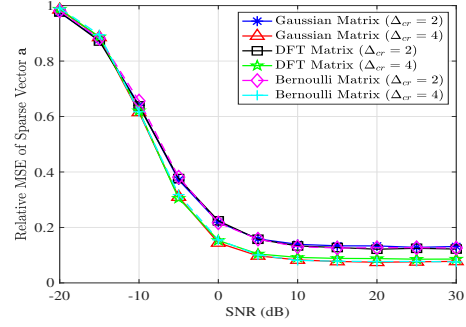} 
\vspace{-0.1cm}
\caption{MSE in recovering $\TarVec$ versus SNR with different compressive measurement matrices (Budget of $2MNL=1728$ bits).}
\label{MSEvsSNR_MeaMtx}
\end{figure}

In Fig. \ref{MSEvsSNR_MeaMtx}, we also compare the estimation performance by using different settings of the matrix $\mathbf{M}$.
Besides the Gaussian matrix, we also show use a matrix with Bernoulli-distributed entries and a \ac{dft} matrix, both of which are widely used in \ac{cs}. 
As shown in Fig. \ref{MSEvsSNR_MeaMtx}, the MSEs in recovering $\TarVec$ with different setting of $\mathbf{M}$ are almost the same, illustrating that the specific choice of  $\mathbf{M}$ hardly affects  the performance of  \ac{bilimo}.

\begin{figure}
\centering
\includegraphics[width=\figWidth, height=\figHeight]{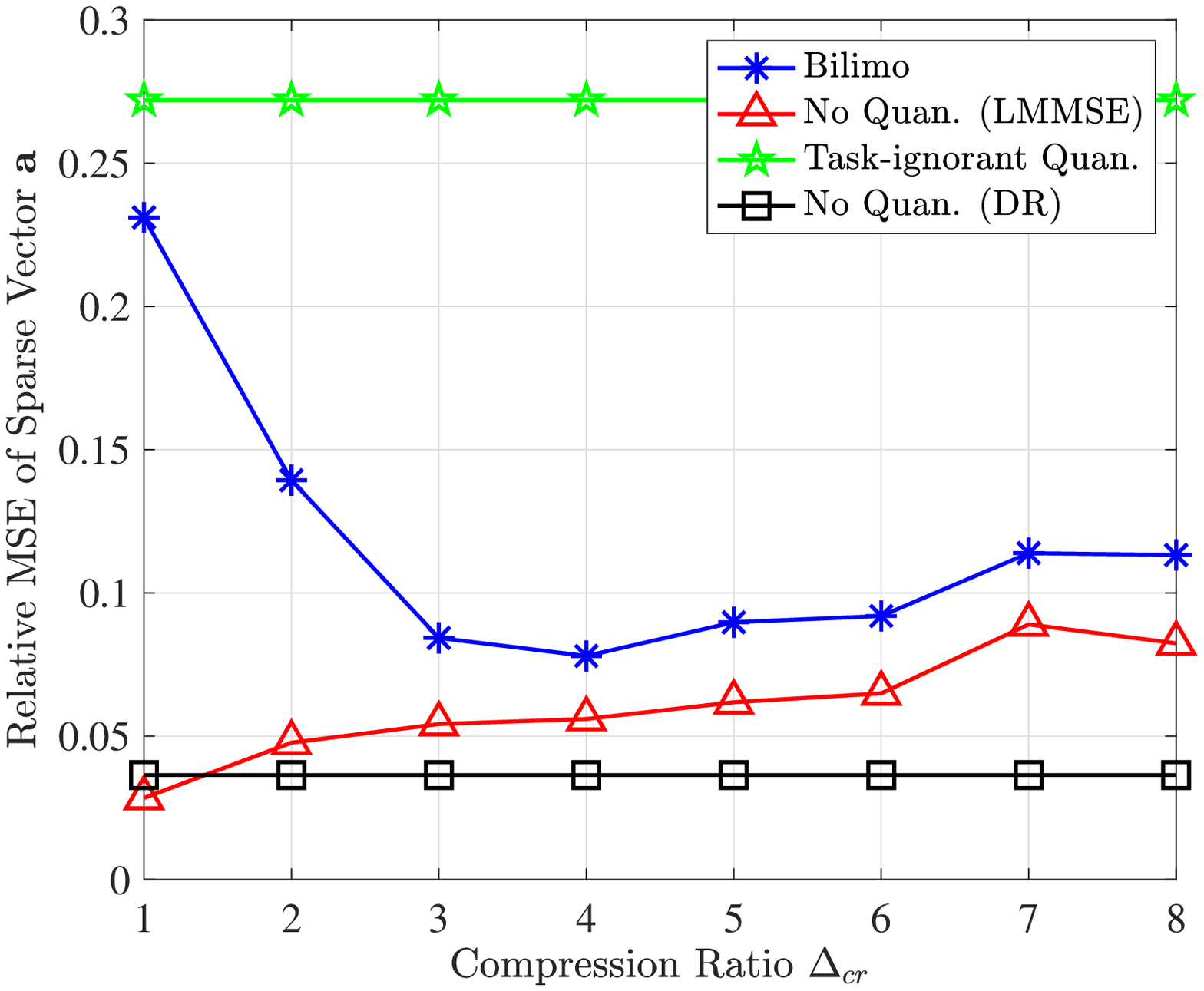} 
\vspace{-0.1cm}
\caption{MSE in recovering $\TarVec$ versus $\Delta_{cr}$ ($2MNL=1728$ bits, SNR=10dB).}
\label{MSEvsCr}
\end{figure}

We conclude our evaluation of the estimation performance of \ac{bilimo} with studying the effect of the compression ratio. To that aim, we evaluate the MSE curves of the \ac{bilimo} receiver and the ``No Quan. (LMMSE)'' method versus the  compression ratio $\Delta_{cr}$, in Fig. \ref{MSEvsCr}.
For comparison, we also depict the MSE curves of the task-ignorant quantization and the ``No Quan. (DR)'' method.
Under the given bit budget, increasing of the compression ratio allows more bits are assigned to each \ac{adc}, at the cost of recovering a further compressed representation of the targets grid. This operation effectively trades sufficiency for quantization distortion, hence for a given overall bit budget we observe a minimum point in which the the compression ratio minimizes the overall MSE performance of the \ac{bilimo} receiver. In particular, for the scenario depicted in Fig. \ref{MSEvsCr}, the \ac{bilimo} receiver achieves the minimal MSE when $\Delta_{cr}=4$, corresponding to 4 bits per ADC. 
The fact that increasing  $\Delta_{cr}$ results in a representation from which sparse recovery induces additional errors  is clearly demonstrated by the MSE curve of the ``No Quan. (LMMSE)'' method.
The results show that the task-based quantization gives rise to a trade-off between the quantization error and the sparse recovery error when setting $\Delta_{cr}$, demonstrating the ability of the \ac{bilimo} receiver to balance these error types.

\vspace{-0.2cm}
\subsection{Detection Performance}
\label{subsec:DetecPerform}
\vspace{-0.1cm}
We  next evaluate the detection accuracy of  \ac{bilimo}, i.e., its ability to detect the positions of the targets encapsulated in the sparsity pattern of   $\mathbf{a}$.
We user the hit rate performance metric, in which a ``hit'' means that the a delay-angle estimate is identical to the true target position.
In our experiments, the amplitude of the reflection coefficients of each target is fixed to unity while its phase   is randomly distributed between $[0,2\pi]$.

\begin{figure}
\centering
\includegraphics[width=\figWidth, height=\figHeight]{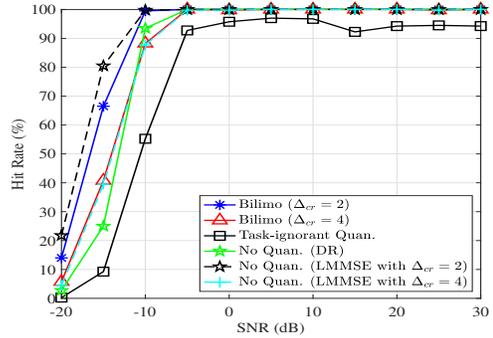} 
\vspace{-0.1cm}
\caption{Hit rate versus SNR ($K = 4$,  budget of $2MNL=1728$ bits).}
\label{HitRatevsSNR}
\end{figure}

We first evaluate the hit rate versus SNR  for a bit budget equivalent to one bit per input sample.
As shown in Fig. \ref{HitRatevsSNR},  \ac{bilimo} outperforms  the task-ignorant quantization, and is within a small gap from the ``No Quan.'' method.
In particular,  \ac{bilimo}  with $\Delta_{cr}=2$ achieves $100\%$ hit rate when the SNR is as low as -10dB, while  task-ignorant quantization does not detect all the target even when the SNR is  30dB due to its dominant quantization error.
Moreover, in low SNRs,  \ac{bilimo}  with $\Delta_{cr}=2$ outperforms the ``No Quan. (DR)'' method.
This  is due to introducing of the  matrix $\mathbf{M}$ used in formulating the compressed vector, which improves the coherence $\mu(\mathbf{M}\mathbf{\Phi})$.

\begin{figure}
\centering
\includegraphics[width=\figWidth, height=\figHeight]{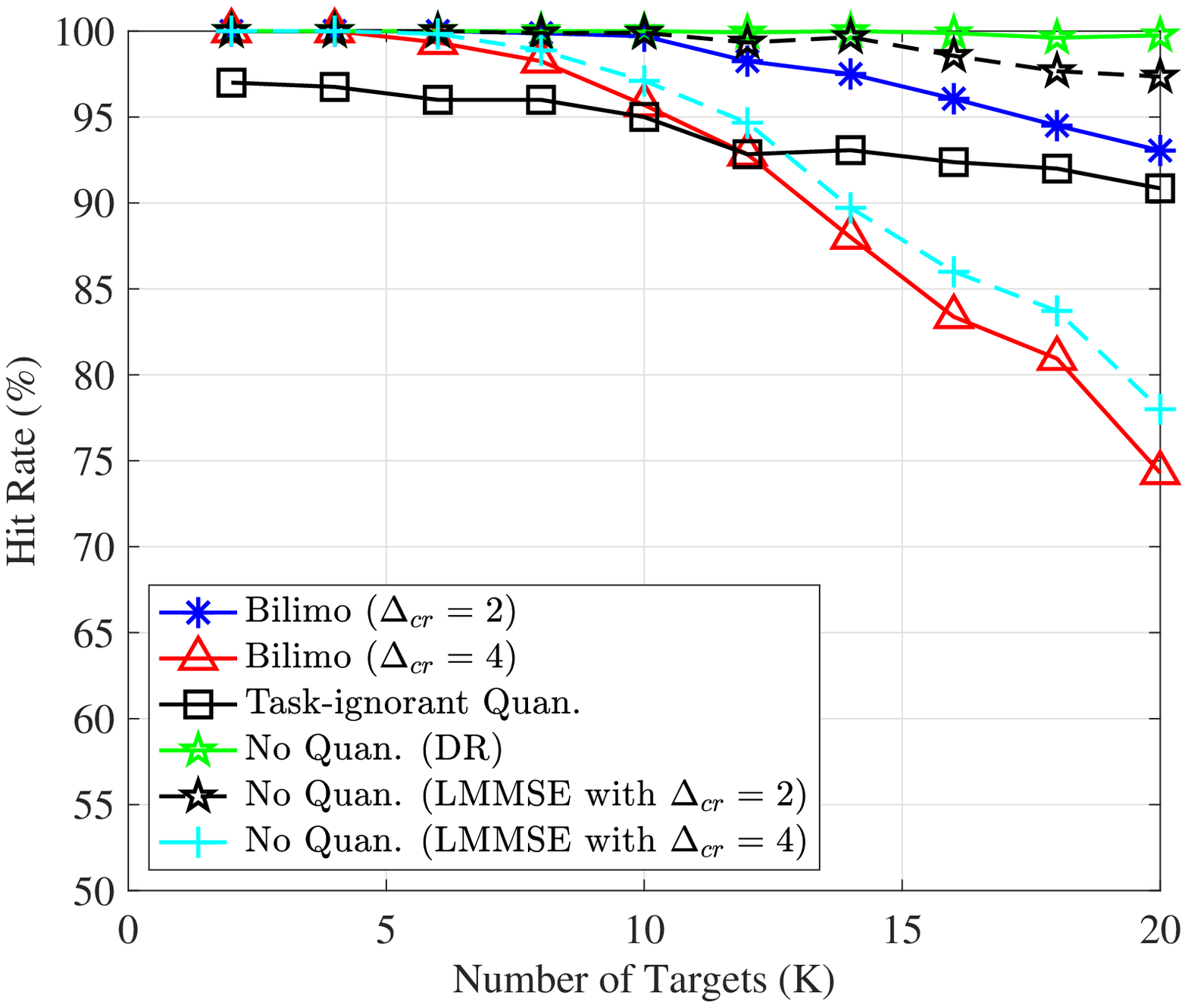} 
\vspace{-0.1cm}
\caption{Hit rate versus  $K$ (SNR=10dB, budget of $2MNL=1728$ bits).}
\label{HitRatevsK}
\end{figure}

Finally, we evaluated the hit rate of various methods versus the number of targets $K$.
The results depicted in Fig. \ref{HitRatevsK} demonstrate that the \ac{bilimo} receiver with $\Delta_{cr}=2$ improves the hit rate over the task-ignorant quantization,
and that the hit rates of all the methods decrease as the number of targets increases.
It is also observed that the detection performance of the \ac{bilimo} receiver is degraded if we increase the compression ratio from $\Delta_{cr}=2$ to $\Delta_{cr}=4$.
This is due to the deterioration of the sparse recovery performance as the compression ratio increases.
Nonetheless, the \ac{bilimo} receiver with $\Delta_{cr}=4$ still outperforms task-ignorant quantizers when $K\leq12$, further demonstrating the benefits of task-based quantization when operating under tight bit constraints.

\vspace{-0.2cm}
\section{Conclusion}
\label{sec:conclusion}
\vspace{-0.1cm}
In this work we designed a bit-limited MIMO radar  receiver, operating with a HAD architecture. We jointly optimized the components of the hybrid receiver, allowing it to  accurately recover the targets while operating with low resolution scalar ADCs.
Our design is built upon the combination of the compressive sensing and the task-based quantization, allowing us to accurately quantify the MSE in recovering the desired target parameters via sparse recovery. Our numerical results demonstrate that the proposed \ac{bilimo} receiver notably outpeforms the conventional approach of recovering the targets solely in the digital domain, and allows to approach the performance achievable with infinite resolution quantizers while operating with a bit budget equivalent to one bit per sample.  

\vspace{-0.2cm}
\begin{appendix}
\numberwithin{proposition}{subsection} 
\numberwithin{lem}{subsection} 
\numberwithin{corollary}{subsection} 
\numberwithin{remark}{subsection} 
\numberwithin{equation}{subsection}	
\vspace{-0.2cm}
\subsection{Proof of Theorem \ref{theorem2}}
\label{app:Proof2} 
Under Assumption A1, we first ﬁnd for each block diagonal matrix $\bar{\mathbf{B}}$ an optimal unitary and block diagonal matrix, which minimizes the $\EMSE(\bar{\mathbf{B}})$ given in \eqref{eqn_EMSE_barB}. Defining $\mathbf{T}_i\triangleq\mathbf{M}_i\mathbf{R}_{c_i}$, the result is stated in the following lemma:

\begin{lem}
For any block diagonal matrix $\bar{\mathbf{B}}=\blkdiag\{\mathbf{B}_1,\cdots,\mathbf{B}_L\}$, we can find an optimal unitary and block diagonal matrix $\bar{\mathbf{U}}_{o}$, such that
\ifFullVersion
\begin{align}
    \EMSE(\bar{\mathbf{B}})\geq &\EMSE(\bar{\mathbf{U}}_{o}\bar{\mathbf{B}})
     =\sum_{i=1}^L\Tr\bigg[ \mathbf{T}_i\bigg(\CovMat^{-1}_i-\mathbf{B}_i^H\big(\mathbf{B}_i\CovMat_i\mathbf{B}_i^H \notag \\ 
     &+\frac{4\eta^2}{3b^2LP}\sum_{i=1}^L\Tr(\mathbf{B}_i\CovMat_i\mathbf{B}_i^H)\mathbf{I}_{P}\big)^{-1}\mathbf{B}_i\bigg)\mathbf{T}^H_i \bigg].
     \label{eqn_EMSEBmin} 
\end{align} 
\else 
$\EMSE(\bar{\mathbf{B}})\geq \EMSE(\bar{\mathbf{U}}_{o}\bar{\mathbf{B}})
     =\sum_{i=1}^L\Tr\Big[ \mathbf{T}_i\Big(\CovMat^{-1}_i-\mathbf{B}_i^H\big(\mathbf{B}_i\CovMat_i\mathbf{B}_i^H  +\frac{4\eta^2}{3b^2LP}\sum_{i=1}^L\Tr(\mathbf{B}_i\CovMat_i\mathbf{B}_i^H)\mathbf{I}_{P}\big)^{-1}\mathbf{B}_i\Big)\mathbf{T}^H_i \Big]$.
\fi
\end{lem}

\begin{IEEEproof}
Note that for any unitary and block diagonal matrix $\bar{\mathbf{U}}$, it follows from (\ref{eqn_EMSE_barB}) that
\ifFullVersion
\begin{align}
        &\EMSE(\bar{\mathbf{U}}\bar{\mathbf{B}})=\Tr\left[\mathbf{T}\CovMat^{-1}\mathbf{T}^H \right]-\Tr\bigg[ \mathbf{T}\bar{\mathbf{B}}^H\Big(\bar{\mathbf{B}}\CovMat\bar{\mathbf{B}}^H \notag \\
        &+\frac{4\eta^2}{3b^2}\max_{i}\{(\bar{\mathbf{F}}\bar{\mathbf{U}}\bar{\mathbf{B}}\CovMat\bar{\mathbf{B}}^H\bar{\mathbf{U}}^H\bar{\mathbf{F}}^H)_{i,i}\}\mathbf{I}_{LP}\Big)^{-1}\!\bar{\mathbf{B}}\mathbf{T}^H \bigg], 
        \label{eqn_EMSEUB} 
\end{align}
\else
$\EMSE(\bar{\mathbf{U}}\bar{\mathbf{B}})=\Tr\left[\mathbf{T}\CovMat^{-1}\mathbf{T}^H \right]-\Tr\Big[ \mathbf{T}\bar{\mathbf{B}}^H\Big(\bar{\mathbf{B}}\CovMat\bar{\mathbf{B}}^H  +\frac{4\eta^2}{3b^2}\max_{i}\{(\bar{\mathbf{F}}\bar{\mathbf{U}}\bar{\mathbf{B}}\CovMat\bar{\mathbf{B}}^H\bar{\mathbf{U}}^H\bar{\mathbf{F}}^H)_{i,i}\}\mathbf{I}_{LP}\Big)^{-1}\!\bar{\mathbf{B}}\mathbf{T}^H \Big]$,
\fi
where $\mathbf{T}=\mathbf{M}\mathbf{P}^T\mathbf{R}_c$. Thus, the optimal unitary and block diagonal matrix $\bar{\mathbf{U}}_{o}$ which minimizes the EMSE is given by
\vspace{-0.1cm}
\begin{equation}\label{prob_barU_o}
\begin{split}
    \bar{\mathbf{U}}_{o}=\arg\min_{\bar{\mathbf{U}}} &\max_{i=1,\cdots,LP}\{(\bar{\mathbf{F}}\bar{\mathbf{U}}\bar{\mathbf{B}}\CovMat\bar{\mathbf{B}}^H\bar{\mathbf{U}}^H\bar{\mathbf{F}}^H)_{i,i}\}\\
    s.t. \quad&\bar{\mathbf{U}}=\blkdiag\{\mathbf{U}_1,\cdots,\mathbf{U}_L\}.
\end{split}    
\vspace{-0.1cm}
\end{equation}
Note that $\bar{\mathbf{F}}$ is also unitary. 
Since for any vector $\mathbf{x}\in\mathbb{C}^L$, $\mathbf{F}_L^H\diag(\mathbf{x})\mathbf{F}_L$ is a circulant matrix with identical diagonal entries, then   $\bar{\mathbf{F}}\bar{\mathbf{B}}\CovMat\bar{\mathbf{B}}^H\bar{\mathbf{F}}^H$ has block structure with identical blocks, where each block has $P$ elements.
Thus, \eqref{prob_barU_o} is equivalent to finding  $\mathbf{U}_1,\cdots,\mathbf{U}_L$ s.t. each block of $\bar{\mathbf{U}}\bar{\mathbf{B}}\CovMat\bar{\mathbf{B}}^H\bar{\mathbf{U}}^H$, i.e.,  $\mathbf{U}_i\mathbf{B}_i\CovMat_i\mathbf{B}_i^H\mathbf{U}_i^H$, has identical diagonal entries.

By Majorization theory  \cite[Cor. 2.4]{Palomar-Major}, it holds that $\min_{\mathbf{U}_i}\max_{i }(\mathbf{U}_i\mathbf{B}_i\CovMat_i\mathbf{B}_i^H\mathbf{U}_i^H)=\frac{1}{P}\Tr(\mathbf{B}_i\CovMat_i\mathbf{B}_i^H)$.
Plugging this into the expression for $\EMSE(\bar{\mathbf{U}}\bar{\mathbf{B}})$ and exploiting the block diagonal structure of $\mathbf{T}$, $\bar{\mathbf{B}}$, and $\CovMat$, proves the lemma.
\end{IEEEproof}

Next, we characterize the block diagonal matrix $\bar{\mathbf{B}}$ which minimizes $\EMSE(\bar{\mathbf{U}}_{o}\bar{\mathbf{B}})$.
Let $\tilde{\mathbf{B}}_i\triangleq\mathbf{B}_i\CovMat^{1/2}_i$, $\LMMSEMatW_i\triangleq\mathbf{T}_i\CovMat^{-1/2}_i$.
\ifFullVersion
Our problem is equivalent to
\begin{align}
\!	\max_{ \{\tilde{\mathbf{B}}_i\}}\sum_{i=1}^L\!\Tr\bigg[ \LMMSEMatW_i\tilde{\mathbf{B}}^H_i&\left(\tilde{\mathbf{B}}_i\tilde{\mathbf{B}}^H_i\!+\!\frac{4\eta^2}{3b^2PL}\sum_{i=1}^L\Tr(\tilde{\mathbf{B}}_i\tilde{\mathbf{B}}^H_i)\mathbf{I}_{P}\right)^{-1} \notag \\ &\times \tilde{\mathbf{B}}_i\LMMSEMatW^H_i\bigg].\label{eqn_maxBlkB}
\end{align}
\else
Our objective is equivalent to
$\sum_{i=1}^L\!\Tr\Big[ \LMMSEMatW_i\tilde{\mathbf{B}}^H_i\Big(\tilde{\mathbf{B}}_i\tilde{\mathbf{B}}^H_i\!+\!\frac{4\eta^2}{3b^2PL}\sum_{i=1}^L\Tr(\tilde{\mathbf{B}}_i\tilde{\mathbf{B}}^H_i)\mathbf{I}_{P}\Big)^{-1}   \tilde{\mathbf{B}}_i\LMMSEMatW^H_i\Big]$. 
\fi
Now the right hand side of the objective is invariant to replacing $\tilde{\mathbf{B}}_i$ with $\alpha\mathbf{U}_i\tilde{\mathbf{B}}_i$ for any $\alpha>0$ and for any unitary $\mathbf{U}_i$.
So we can fix $\Tr(\tilde{\mathbf{B}}_i\tilde{\mathbf{B}}^H_i)=1$, and write $\tilde{\mathbf{B}}_i=\boldsymbol\Lambda_i\mathbf{V}^H_i$, where
$\boldsymbol\Lambda_i\in\mathbb{R}^{P\times MN}$ is a diagonal matrix whose  entries are arranged in a descending order, and $\mathbf{V}_i $ is unitary.
Our objective now becomes
\ifFullVersion
\begin{align} 
	\max_{\{\boldsymbol\Lambda_i,\mathbf{V}_i\}}&\sum_{i=1}^L\Tr\left[ \LMMSEMatW^H_i\LMMSEMatW_i\mathbf{V}_i\boldsymbol\Lambda^T_i\left(\boldsymbol\Lambda_i\boldsymbol\Lambda^T_i\!+\!\frac{4\eta^2}{3b^2P}\mathbf{I}_{P}\right)^{-1}\!\!\boldsymbol\Lambda_i\mathbf{V}^H_i\right] \notag\\
	s.t. \quad &\Tr(\boldsymbol\Lambda_i\boldsymbol\Lambda^T_i)=1, i=1,2,\cdots,L.
\label{eqn_maxBlkB2}
\end{align}
Solving \eqref{eqn_maxBlkB2}
\else
$\sum_{i=1}^L\Tr\Big[ \LMMSEMatW^H_i\LMMSEMatW_i\mathbf{V}_i\boldsymbol\Lambda^T_i\left(\boldsymbol\Lambda_i\boldsymbol\Lambda^T_i\!+\!\frac{4\eta^2}{3b^2P}\mathbf{I}_{P}\right)^{-1}\!\!\boldsymbol\Lambda_i\mathbf{V}^H_i\Big]$, subject to $\Tr(\boldsymbol\Lambda_i\boldsymbol\Lambda^T_i)=1$ for each $i$. This 
\fi
is the same as solving 
\begin{align*}
	\!\max_{\boldsymbol\Lambda_i,\mathbf{V}_i}\!\Tr\Big[ \LMMSEMatW^H_i\LMMSEMatW_i\mathbf{V}_i\boldsymbol\Lambda^T_i\left(\boldsymbol\Lambda_i\boldsymbol\Lambda^T_i\!+\!\frac{4\eta^2}{3b^2P}\mathbf{I}_{P}\right)^{-1}\!\!\boldsymbol\Lambda_i\mathbf{V}^H_i\Big],
\end{align*}
 for each $i$, subject to $\Tr(\boldsymbol\Lambda_i\boldsymbol\Lambda^T_i)=1$,   
\ifproofs
\else 
which is equivalent to \cite[Eq. (C.8)]{Shlezinger-TSP2019}.
Hence, following the derivation    in \cite[Appendix C]{Shlezinger-TSP2019} proves the theorem.
\fi
\qed
\end{appendix}
 
\bibliographystyle{IEEEtran}
\bibliography{IEEEabrv,MIMO,Onebit,TaskQuan}

\end{document}